\documentclass[%
 reprint,
 superscriptaddress,
 amsmath,amssymb,
]{revtex4-1}

\usepackage{graphicx}
\usepackage{dcolumn}
\usepackage{bm}
\usepackage{makecell}
\usepackage{epsfig}
\usepackage{subcaption}
\usepackage{verbatim}
\usepackage{setspace}
\usepackage{xr}
\graphicspath{{../figs/}}
\usepackage[mathlines]{lineno}

\begin{document}


\title{Snow topography on undeformed Arctic sea ice captured by an idealized ``snow dune'' model}
\author{Predrag Popovi\'{c}}
 \altaffiliation{All correspondence should be directed to Predrag Popovi\'{c}}
  \email{arpedjo@gmail.com}
  \affiliation{%
 The University of Chicago, The Department of the Geophysical Sciences\\
}%
\author{Justin Finkel}%
\affiliation{%
The University of Chicago, Committee on Computational and Applied Mathematics\\
}%
\author{Mary C. Silber}%
\affiliation{%
The University of Chicago, Department of Statistics and Committee on Computational and Applied Mathematics\\
}%
\author{Dorian S. Abbot}
\affiliation{%
The University of Chicago, The Department of the Geophysical Sciences\\
}%




\date{\today}

\begin{abstract}

Our ability to predict the future of Arctic sea ice is limited by ice's sensitivity to detailed surface conditions such as the distribution of snow and melt ponds. Snow on top of the ice decreases ice's thermal conductivity, increases its reflectivity (albedo), and provides a source of meltwater for melt ponds during summer that decrease the ice's albedo. In this paper, \linelabel{Resub.19}we develop a simple model of pre-melt snow topography that accurately describes snow cover of flat, undeformed Arctic sea ice on several study sites for which data was available. The model considers a surface that is a sum of randomly sized and placed ``snow dunes'' represented as Gaussian mounds. This model generalizes the ``void model'' of \citet{popovic2018simple} and, as such, accurately describes the statistics of melt pond geometry. We test this model against detailed LiDAR measurements of the pre-melt snow topography. We show that the model snow-depth distribution is statistically indistinguishable from the measurements on flat ice, while small disagreement exists if the ice is deformed. We then use this model to determine analytic expressions for the conductive heat flux through the ice and for melt pond coverage evolution during an early stage of pond formation. We also formulate a criterion for ice to remain pond-free throughout the summer. Results from our model could be directly included in large-scale models, thereby improving our understanding of energy balance on sea ice and allowing for more reliable predictions of Arctic sea ice in a future climate. 

\end{abstract}

\maketitle

\section{Introduction}

As a hallmark of climate change and a major component of the Arctic environment, Arctic sea ice retreat has garnered much scientific and media attention \citep{perovich2009loss,stroeve2007arctic}. Predicting the rate of sea ice loss is a challenge. The difficulty lies in the fact that sea ice evolution is controlled by many processes that operate on scales ranging from sub-millimeter to tens of kilometers \citep{holland1999role}. Possibly the most notable among these processes is the interaction of ice with fluxes of energy coming from the environment mediated by detailed conditions on the ice surface. The presence of snow, water, dirt, \linelabel{Resub.41}or black carbon on the ice surface can drastically change the rate of absorption of solar radiation or the rate of thermal conduction during winter growth \citep{perovich1996optical}. Finding ways to reduce the complexity of modeling ice surface conditions is of great importance for accurately determining the ice energy balance in large-scale models and, therefore, for improving our understanding of the future of sea ice in a changing climate.

Snow insulates the ice \citep{yen1981review,sturm2002thermal}, reflects sunlight \citep{perovich1996optical}, and provides a source of fresh water that collects into melt ponds \citep{polashenski2017percolation}. The net effect of each of these processes depends on the spatial distribution of the snow cover. \linelabel{ResubResub.67}This is important because snow often exhibits significant spatial variability that arises from a multitude of bedforms, most notable of which are meter-scale snow dunes \citep{mather1962further,filhol2015snow}. Ice covered with patchy snow will, on average, grow faster than if the same amount of snow were spread uniformly, since uniform cover insulates the ice more thoroughly \citep{sturm2002thermal}. Second, uniform snow cover also protects the underlying ice more from solar radiation than a patchy layer, since even a thin layer of snow can increase albedo significantly \citep{perovich1996optical}. Finally, as snow starts to melt, melt ponds first appear in the regions where there is the least snow \citep{petrich2012snow}. For this reason, patchy snow cover leads to melt ponds covering the ice surface sooner, melting more ice. In summary, patchy snow cover will lead to both more ice growth during winter as well as to more ice melt during summer. 

\citet{liston2018distributed} modeled snow on flat ice using a simple statistical approach. They represented the snow surface by smoothing and rescaling an initially random height field to match the measured mean snow depth, its variance, and the horizontal correlation length. Their approach is likely sufficient for many practical applications where snow topography needs to be included. However, it produces some essentially unphysical predictions, such as negative snow depths in cases when snow depth standard deviation is comparable to the mean. Here, we adopt a somewhat similar approach that is physically consistent and reproduces the measurements with high accuracy. 

As we mentioned above, snow often exists in meter-scale dunes which are the most prominent features of the snow cover \citep{mather1962further,filhol2015snow}. In \citet{popovic2018simple}, we considered a simple geometric ``void model'' which represented these snow dunes as circles, and compared it to aerial photographs of melt ponds. These circles had varying size and were placed on a surface randomly and independently of each other, while melt ponds were represented as voids between these circles. Our model was able to reproduce the geometric statistics of melt ponds, such as their size distribution and the fractal dimension as a function of pond size, over the entire observational range of more than 6 orders of magnitude in pond area. 

In this paper, we generalize the two-dimensional discontinuous ``void model'' to a continuous synthetic ``snow dune'' topography that has a vertical component. We represent snow dunes as mounds of Gaussian form that have a randomly chosen horizontal scale, a height proportional to that scale, and that are placed randomly on the surface. The surface topography is the sum of many such mounds. Like the ``void model,'' this topography accurately describes the horizontal melt pond features, which we take as indirect evidence that it accurately describes the horizontal snow features. We support this by showing that our model fits both LiDAR scans of the pre-melt snow topography and melt pond data using the same typical horizontal mound scale. The fact that the horizontal scale is so similar is somewhat surprising since the two datasets were recoded in different years, different times of year, and different ice types. 

The novel prediction of our continuous topography is a vertical snow-depth distribution. By comparing moments of our model distribution to LiDAR measurements of snow-depth, we show that our model height distribution is indistinguishable from the measurements for two occasions when the underlying ice was very flat. The one LiDAR measurement on slightly deformed ice showed subtle deviation from our model. Thus, we hypothesize that our model is a highly accurate representation of snow on undeformed ice, while it becomes increasingly inaccurate on deformed ice. The close agreement between snow topography and our model is the main result of this paper. Since our model has only three parameters that are uniquely determined by the mean, variance, and the correlation length of snow depth, it appears that it is only necessary to measure these three quantities to completely characterize the statistics of snow cover on flat, undeformed ice. 

After showing the agreement between our model and snow measurements, we consider some applications of our model. First, we consider how spatial variability in snow affects the heat conduction through the ice during winter ice growth. \citet{sturm2002thermal} conducted measurements of snow conductivity, $k_s$, on deformed multi-year ice, and found that $k_s$ inferred from ice growth is $\sim 2.4$ times higher than $k_s$ inferred from direct measurements. They found that about 40\% of the increase can be explained by snow and ice geometry and that horizontal heat transport, typically neglected in large-scale models, likely contributes significantly. To understand whether these conclusions also hold for undeformed ice, we solve a 3-dimensional heat equation within the ice assuming that the snow cover is well-described by our ``snow dune'' model. We develop a simple analytic equation to determine the heat flux conducted through the ice given a set of physical parameters. \linelabel{ResubResub.110}Contrary to \citet{sturm2002thermal}, our model shows a negligible horizontal heat transport. Moreover, our model shows a smaller increase in heat conduction than \citet{sturm2002thermal}. This disagreement could be due to the fact that our model assumes undeformed ice while the field measurements of \citet{sturm2002thermal} were made on deformed ice.

Next, we consider the effect of snow topography on early melt pond development. Using the fact that our model height distribution is well-fit with a gamma distribution, we write an analytic evolution equation for melt pond coverage during an early stage when ice is impermeable. This equation for melt pond evolution enables us to understand how early-stage melt pond growth depends on measurable environmental parameters. Moreover, it allows us to derive a simple condition that distinguishes whether ice will remain pond-free throughout the summer based on snow depth, variance, density and melt rate. 

This paper is organized so that the new results are presented in the main text, while confirming previous results about melt ponds on our topography and mathematical details are left for the Supplementary Information (SI). In section \ref{sec:topo}, we describe our model of the ``snow dune'' topography. Next, in section \ref{sec:analytic_stats} we describe the analytical properties of our model such as the dependence of moments and the correlation function on model parameters. Then, in section \ref{sec:snow}, we compare our model with the measured snow topography. In section \ref{sec:heat}, we apply our ``snow dune'' model to determine the conductive heat flux through the ice. Next, in section \ref{sec:stageI}, we use our model to investigate melt pond evolution on flat impermeable ice. Finally, in section \ref{sec:discussion}, we discuss the implications of our results for large-scale studies and in section and \ref{sec:conclusions}, we conclude. In SI section S1, we confirm that the synthetic ``snow dune'' topography predicts ponds that accurately reproduce the geometric statistics of real ponds during late summer. \linelabel{ResubResub.132}In the subsequent SI sections, we prove the mathematical results and show that our model predictions are robust against details such as the shape of the mounds or mound radius.

\section{The synthetic ``snow dune'' topography}\label{sec:topo}

\begin{figure*}
\centering
\includegraphics[width=0.7\linewidth]{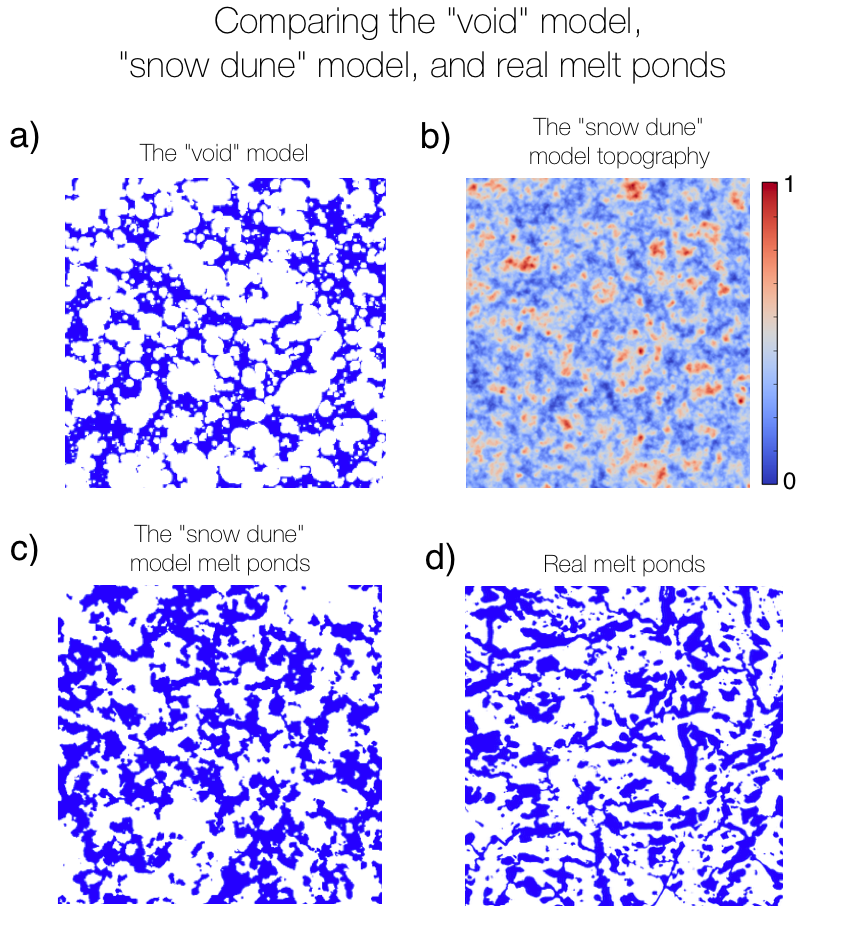}
\caption{a) A realization of the ``void model.'' Figure is taken from \citet{popovic2018simple}. b) Synthetic ``snow dune'' topography. Red colors indicate topographic highs, while blue colors indicate topographic lows. The upper bound on the scale bar, here set to 1, is arbitrary. c) Ponds on a ``snow dune'' topography. d) A binarized image of a real melt pond photograph taken on August 14th during the HOTRAX mission.}
\label{fig:topos}
\end{figure*}

In this section we will describe our synthetic ``snow dune'' topography. The entire rest of the paper hinges on this topography.

In \citet{popovic2018simple}, we developed a simple geometric model where snow dunes were represented by overlapping circles of varying size placed randomly and independently of each other on the ice surface, while melt ponds were represented as voids between these circles. We then compared the statistical properties associated with this simple geometric model with observations of late-summer melt ponds. We found a remarkable agreement between various statistics of real and model melt ponds: by tuning only two model parameters, the model matched the measured pond area distribution, the pond fractal dimension as a function of pond area, and two correlation functions that describe pond size and connectedness, over the entire observational range of nearly 7 orders of magnitude in pond area. A realization of this model is shown in Fig. \ref{fig:topos}a. 

\linelabel{Resub.145}Motivated by the success of the ``void model'' in capturing the conditions on the ice surface, we set about to generalize it to a 3-dimensional ``snow dune'' topography that has an additional vertical component, and can, thus, be compared with a pond-free ice surface. We will show that, in addition to reproducing melt pond features, the ``snow dune'' model also matches the observed snow-depth distribution and correlation function on flat, undeformed ice.

To generalize the ``void model'' to a continuous topography, we replace circles with mounds that have a vertical profile. The mounds have a Gaussian shape
\begin{linenomath*}
\begin{equation}\label{mound1}
h(\mathbf{x}) = h_m e^{-(\mathbf{x}-\mathbf{x}_0)^2 / 2 r^2} \text{  , }
\end{equation}
\end{linenomath*}
where $h(\mathbf{x})$ is the height of the mound at a point $\mathbf{x}$ on the surface, $\mathbf{x}_0$ is the location of the center of the mound, $r$ is the horizontal scale, and $h_m$ is the maximum height of the mound. Mounds are placed randomly, i.e., $\mathbf{x}_0$ can be anywhere on the surface with equal probability. The horizontal scale, $r$, is randomly chosen from an exponential probability distribution, $f_r$ 
\begin{linenomath*}
\begin{equation}\label{frofr}
f_r(r) = \frac{1}{r_0}e^{-r/r_0} \text{  , } 
\end{equation}
\end{linenomath*}
where $r_0$ is the typical scale of the mounds. \linelabel{Resub.156B}This distribution is the same as the distribution of circle radii in the ``void model'' of \citet{popovic2018simple}, and we chose it mainly due to its simple form and the fact that it depends on a single parameter, $r_0$. Otherwise, the choice of this distribution is arbitrary and does not affect the main conclusions of our model so long as it is not long-tailed. To prevent having unrealistically narrow and high mounds, we prescribed the height of each mound, $h_m$, to be proportional to its horizontal scale, $h_m = h_{m,0} r/r_0$, where $h_{m,0}$ is the typical mound height. The topography is then a sum of $N$ such mounds placed on an initially flat surface 
\begin{linenomath*}
\begin{equation}
h_\text{SD}(\mathbf{x}) = \sum_{i=1}^{N} h_{m,0} \frac{r_i}{r_0} e^{-(\mathbf{x}-\mathbf{x}_{0_i})^2 / 2 r_i^2} \text{  , }
\end{equation}
\end{linenomath*}
where $h_\text{SD}(\mathbf{x})$ is the height of the ``snow dune'' topography at location $\mathbf{x}$. A realization of this topography is shown in Fig. \ref{fig:topos}b. Ponds that form on this surface after cutting it with a horizontal plane are shown in Fig. \ref{fig:topos}c. This is shown alongside a realization of the ``void model'' (Fig. \ref{fig:topos}a) and a binarized image of real melt ponds (Fig. \ref{fig:topos}d).

There are three parameters in this model. These are the typical horizontal mound scale, $r_0$, the density of mounds, $\rho$, placed within the domain of size $L$, $\rho \equiv N \frac{r_0^2}{L^2}$, and the typical mound height, $h_{m,0}$. The first two of these parameters, $r_0$ and $\rho$, also enter the ``void model.''

\linelabel{Resub.155}\linelabel{Resub.M1.1}Although our model mounds bare some resemblance with real snow dunes, our model is not accurate on the scale of individual dunes. In particular, snow dunes are certainly not Gaussian-shaped, they show strong anisotropy, \linelabel{Resub.156}the distribution of dune sizes is likely not exponential, and dunes sometimes form semi-regular lattices, which violates the random placement assumption \citep{kochanski2019evolution,filhol2015snow,sharma2019understanding,kochanski2019rescal}. In SI subsection S2.3, we show that aggregate, radially averaged statistical properties of the snow surface that we consider here do not depend on most of these details. We also note that the random placement assumption is important for the conclusions we make, so our model can be useful for large scale studies only if snow surfaces that strongly violate this assumption do not occur often in nature.

\section{Statistics of the synthetic ``snow dune'' topography}\label{sec:analytic_stats}

Here, we describe the statistics of the ``snow dune'' model that we will use later. In particular, we first show how the mean, variance, and correlation length depend on model parameters. We then derive all of the moments of the topography. Finally, we compare the height distribution of the ``snow dune'' topography with a gamma distribution. We derive all of the results in this section in SI section S2. All of these results follow directly from the definition of the ``snow dune'' model we described in section \ref{sec:topo}.   

First, we find that the mean, $\langle h_\text{SD}\rangle$, and variance, $\sigma^2(h_\text{SD}) = \langle (h_\text{SD})^2 \rangle - \langle h_\text{SD}\rangle^2$, of the topography depend on the typical mound height, $h_{m,0}$, and the density of mounds, $\rho$, as
\begin{linenomath*}
\begin{equation}\label{snowdunestats}
\langle h_\text{SD} \rangle = 12 \pi h_{m,0} \rho
	\quad\text{ , }\quad  
\sigma ^2(h_\text{SD}) = 24 \pi h_{m,0}^2 \rho \text{  . }
\end{equation}
\end{linenomath*}
In fact, we can explicitly find every moment of the topography, $\big\langle (h_\text{SD})^n \big\rangle$, by knowing the previous moments and using the recursion formula (see SI section S2 for derivation)
\begin{linenomath*}
\begin{equation}\label{momentsmain}
\big\langle (h_\text{SD})^n \big\rangle = 2 \pi \rho \sum_{j=1}^n  {n-1\choose n-j} \frac{(j+2)!}{j} h_{m,0}^j  \big\langle (h_\text{SD})^{n-j} \big\rangle \text{  . }
\end{equation}
\end{linenomath*}
Equations \ref{snowdunestats} also follow from this formula. 

Next, we define the height correlation function, $C_h(\mathbf{l})$, as 
\begin{linenomath*}
\begin{equation}
 C_h(\mathbf{l}) \equiv  \frac{\langle h_\text{SD}(\mathbf{x})h_\text{SD}(\mathbf{x} + \mathbf{l}) \rangle - \langle h_\text{SD}\rangle^2}{ \sigma ^2(h_\text{SD}) } \text{  , }  \label{autocorr} 
\end{equation}
\end{linenomath*}
where $\mathbf{l}$ is a horizontal displacement of the topography. \linelabel{ResubResub.209}Below, we will use the boldface notation, $\mathbf{l}$, to denote the vector displacement and we will use the italic notation, $l$, to denote the magnitude of this displacement. The correlation function quantifies how much the surface height at $\mathbf{x}$ is correlated with the surface height at $\mathbf{x} + \mathbf{l}$. If the surface is isotropic, $C_h$ only depends on the magnitude of the displacement vector, $l$. We can then define the correlation length, $l_0$, as the distance at which $C_h$ falls by a factor of $e$. Beyond $l_0$, features on the topography can be considered approximately uncorrelated. Using these definitions, we find that $C_h$ and $l_0$ depend on the horizontal mound scale, $r_0$, of an isotropic ``snow dune'' model as (see SI section S2 for derivation)
\begin{linenomath*}
\begin{align}
C_h(l) & = \frac{1}{ 24 } \int_0^\infty z^4 e^{-z-l^2/(2r_0z)^2} \text{d} z  \text{  , }  \label{autocorr_exact} \\
l_0 &= r_0 \xi_0  \text{  , } \label{corr_length}
\end{align}
\end{linenomath*}
where $\xi_0 \approx 9.3689\text{...}$ is a number that can be estimated to arbitrary precision by inverting $C_h(l_0) = 1/e$ using Eq. \ref{autocorr_exact}. The mean, variance, and the correlation length are all measurable properties of real topographies. \linelabel{ResubResub.216}This means that computing these quantities from measurements lets us unambiguously choose the model parameters limited only by the accuracy of those measurements.

\begin{figure*}
\centering
\includegraphics[width=0.6\linewidth]{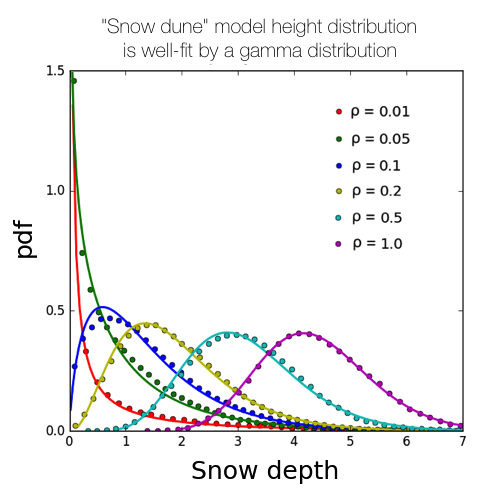}
\caption{Dots represent the height distribution of the ``snow dune'' topography for different densities of mounds, $\rho$. Solid lines represent fits using a gamma distribution. Units of snow depth here are arbitrary. Note that the LiDAR measurements we consider in section \ref{sec:snow} range from $\rho \approx 0.2$ to $\rho \approx 0.5$, a much smaller range than we investigated here. }
\label{fig:snowdunegamma}
\end{figure*}

Since we can find all of the moments of the ``snow dune'' topography using Eq. \ref{momentsmain}, we can fully determine its height distribution. However, it is impractical to do this, and, for practical purposes, we now show that the height distribution of the ``snow dune'' topography is well-fit with a gamma distribution. The two distributions are not the same, as can be seen by comparing their moments (see SI figure S2). However, they are qualitatively very similar, and the differences between them arise from arbitrary choices in our model such as the exponential distribution of radii of the snow dunes, Eq. \ref{frofr}, and the linear relationship between the mound radius and height. We describe the relationship between the height distribution of the ``snow dune'' topography and the gamma distribution in detail in SI section S2.

A gamma distribution has the form
\begin{linenomath*}
\begin{equation} \label{gamma_dist}
f_\Gamma (h) = \frac{1}{\Gamma (k) h_0^k} h^{k-1} e^{-h/h_0} \text{  ,  }
\end{equation}
where $h_0$ is a scale parameter, $k$ is a shape parameter, and $\Gamma (x)$ is a gamma function. The two parameters of the gamma distribution, $h_0$ and $k$, are related in a simple way to the mean and variance of the distribution, $\langle h \rangle$ and $\sigma ^2(h)$
\begin{equation}\label{gammaparams}
k = \frac{\langle h \rangle^2}{\sigma ^2(h)} 
	\quad\text{ , }\quad  
h_0 = \frac{\sigma ^2(h)}{\langle h \rangle} \text{  . }
\end{equation}
\end{linenomath*}
Both of these parameters are therefore uniquely determined by the typical height and density of mounds through Eqs. \ref{snowdunestats}. In particular, we find that the scale parameter, $h_0$, scales linearly with the typical height of mounds, while the shape parameter, $k$, scales linearly with the density of mounds,
\begin{linenomath*}
\begin{equation}\label{kandh0}
h_0 = 2 h_{m,0} \quad \text{ , } \quad k = 6 \pi \rho \text{  . }
\end{equation}
\end{linenomath*}

In Fig. \ref{fig:snowdunegamma}, we show height distributions for synthetic topographies with varying spatial densities of mounds, $\rho$, with the variance, $\sigma^2(h_\text{SD})$, kept fixed. We can see that varying $\rho$ changes the shape of the distribution and that a gamma distribution can fit the height distribution well for all choices of $\rho$. To quantify the quality of the fit, we use the Kolmogorov-Smirnov (KS) statistic, i.e., the maximum distance between the empirical and theoretical cumulative distributions. The fit is typically considered good if this statistic is below 0.05. We find that for all choices of $\rho$, the maximum distance between the cumulative ``snow dune'' height distribution and the best-fit gamma distribution is below 0.05, and ranges between 0.03 for $\rho = 0.01$ to 0.006 for $\rho = 1$, generally decreasing for increasing $\rho$. \linelabel{Resub.223}Our ``snow dune'' model height distribution compares similarly well with a gamma distribution according to other metrics such as the Kuiper or the Cram\'{e}r-von Mises metrics.

\linelabel{Resub.M1}In SI subsection S2.3, we compare different variants of our model that include anisotropy, different dune shapes, and different distributions of mound sizes. There, we show that moments of the height distribution as well as the radially averaged correlation function are insensitive to these details. \linelabel{Resub.164B}Namely, these alterations were only able to change the numerical pre-factors in equations that relate model parameters to measurable quantities, such as $12 \pi$ and $24\pi$ in Eqs. \ref{snowdunestats} or $\xi_0$ in Eq. \ref{corr_length}. Since our conclusions regarding heat conduction and melt pond evolution that we discuss later in the paper only rely on these invariant statistics, they should also be robust to these. We caution, however, that  certain phenomena, such as the wind stress, may explicitly depend on specifics such as snow anisotropy. In such a case, additional parameters in the model describing these details would be required.

\section{Measured snow topography}\label{sec:snow}

\begin{figure*}
\centering
\includegraphics[width=1.\linewidth]{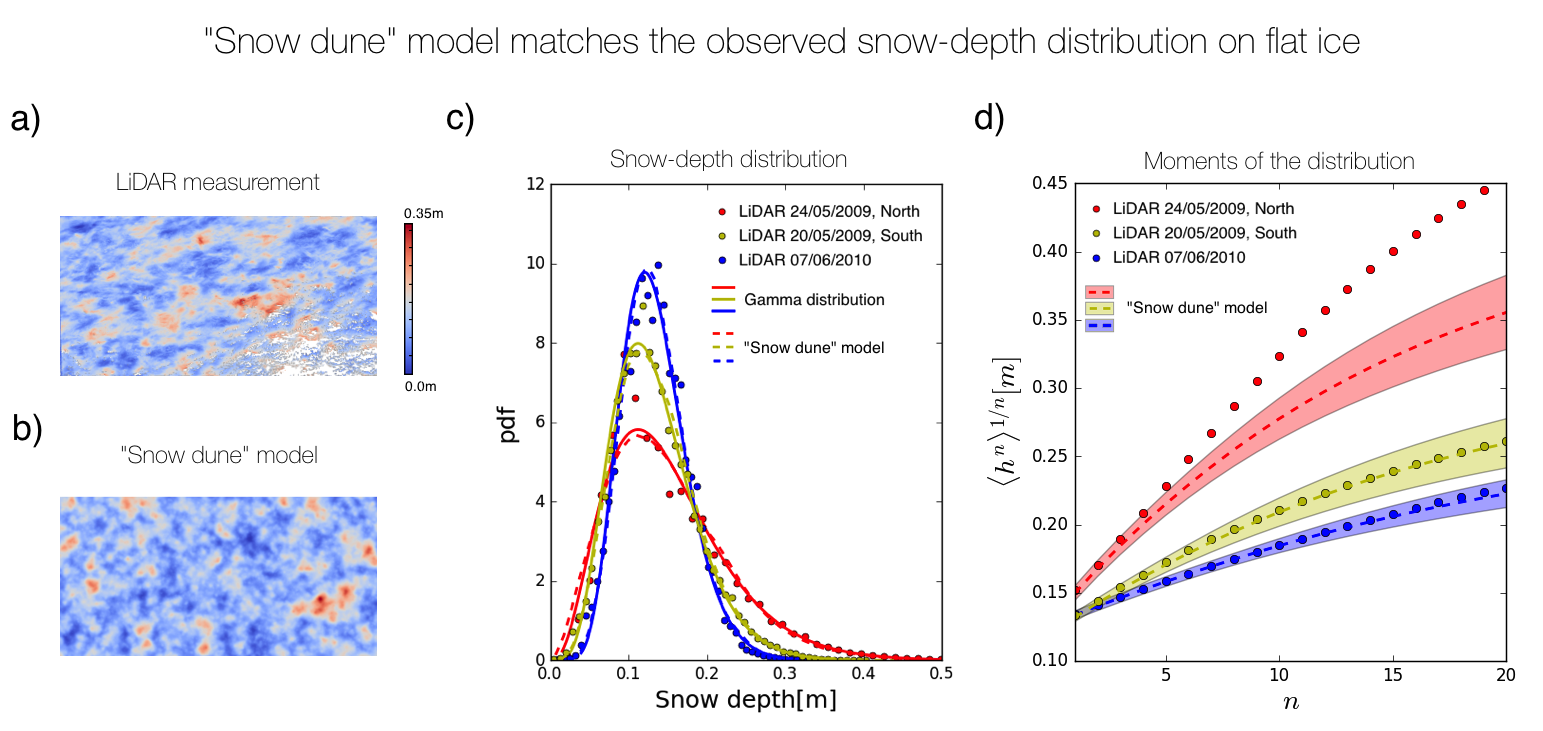}
\caption{a) LiDAR measurement of the pre-melt snow topography by \citet{polashenski2012mechanisms} in early June of 2010. b) A realization of synthetic ``snow dune'' topography. The density of mounds was chosen such that the height distribution of the synthetic topography corresponds well to the measured snow-depth distribution. c) Dots represent snow-depth distribution estimated using LiDAR measurements for two different years. Solid lines are gamma distributions with parameters estimated based on the mean and variance of the measurements. Dashed lines are the corresponding ``snow dune'' topography height distributions, and are nearly indistinguishable from the solid lines. d) Moments of the height distribution. The horizontal axis is the moment order, $n$, while the vertical axis is the $n$-th root of the $n$-th moment. Dots represent moments of the measured snow topography. The thick dashed lines represent the mean moments across an ensemble of 50 randomly-generated ``snow dune'' topographies on the domain of the same size as the measurements. The colored shadings represent one standard deviation from the mean on that ensemble. }
\label{fig:polashsnow}
\end{figure*}

\noindent

The synthetic ``snow dune'' topography is meant to represent the height of snow relative to the flat underlying ice. Therefore, the height distribution of the ``snow dune'' topography should correspond to the pre-melt snow-depth distribution.  Here, we will compare the statistics of the synthetic ``snow dune'' topography to detailed LiDAR measurements of pre-melt snow topography made by \citet{polashenski2012mechanisms}.

\begin{figure*}
\centering
\includegraphics[width=0.8\linewidth]{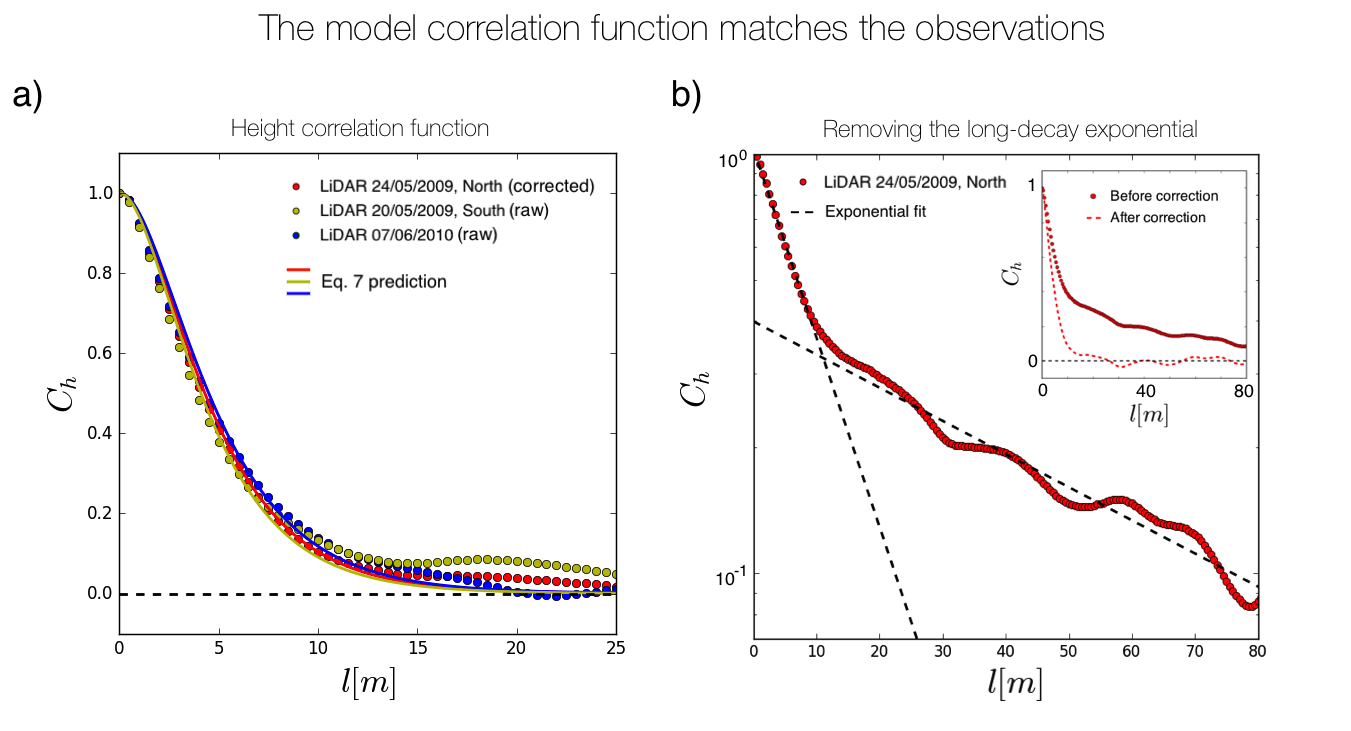}
\caption{a) Height correlation functions. Dots represent measurements, and the solid lines represent predictions for the ``snow dune'' model using Eq. \ref{autocorr_exact}. Correlation functions for 2010 and 2009S measurements (yellow and blue dots) were calculated from raw data. Correlation function for the 2009N measurement was corrected by removing a long-decay exponential. The raw 2009N data are shown in panel b. b) Correction for the 2009N correlation function. Dots represent $C_h$ calculated with raw data on a semi-log plot. The two black dashed lines are exponential fits to the short-decay and long-decay portions of $C_h$. Inset shows $C_h$ before (red dots) and after (red dashed line) removing the long-decay exponential on a linear plot. }
\label{fig:polashcorr}
\end{figure*}

During their field expedition, \citet{polashenski2012mechanisms} performed detailed LiDAR scans of the surface topography within a 100m$\times$200m region on multiple dates in 2009 and 2010 \linelabel{Resub.246}near Barrow, Alaska. During 2009, \citet{polashenski2012mechanisms} monitored two locations separated by about 1 km from each other, one in the north (2009N) and one in the south (2009S). In 2010 they monitored only one location. In Fig. \ref{fig:polashsnow}a we show an example of such a measurement, which shows the height of snow before the start of the melt season, and compare it to a randomly generated ``snow dune'' topography (Fig. \ref{fig:polashsnow}b). Assuming that the underlying ice is flat and the pre-melt ice is fully covered with snow, the LiDAR-estimated height in Fig. \ref{fig:polashsnow}a represents the snow depth plus some reference height. Therefore, to compare with the ``snow dune'' topography, we estimated the snow depth from these LiDAR scans by subtracting the minimum LiDAR elevation from the rest of the scan. We show the snow-depth distribution for the three LiDAR scans in Fig. \ref{fig:polashsnow}c and we show the moments of these distributions in Fig. \ref{fig:polashsnow}d. We estimate the height correlation function for these three measurements in Fig. \ref{fig:polashcorr}. The data from \citet{polashenski2012mechanisms} are freely available at \url{http://chrispolashenski.com/data.php}. 

To see whether the snow-depth distribution of these measurements conforms to the predictions of our model and a gamma distribution, we inferred parameters $h_{m,0}$ and $\rho$ using Eqs. \ref{snowdunestats} and parameters of the gamma distribution, $k$ and $h_0$, using Eqs. \ref{gammaparams} based only on measured mean snow depth and snow depth variance. We show the statistics of measured topographies along with model and gamma distribution parameters in Table \ref{tab:1}. The solid lines in Fig. \ref{fig:polashsnow}c show a gamma distribution with parameters chosen in this way, while the dashed lines show a height distribution of the synthetic ``snow dune'' topography. We can see that in all cases, the measured snow-depth distribution agrees well with our model and a gamma distribution, with a Kolmogorov-Smirnov (KS) statistic remaining at or below 0.03 in all cases. 

\begin{table}
\centering
\setstretch{0.6}
\small
\caption{Statistics of the LiDAR measurements and model parameters inferred from them. The KS statistic was calculated with respect to the gamma distribution.}
\begin{tabular}{| c | c | c | c | c | c | c | c | c | c |}
\hline
& $\langle h \rangle$ & $\sigma(h)$ & $l_0$ & $h_{m,0}$ & $\rho$ & $r_0$ & $h_0$ & $k$ & KS\\
\hline		
2009N & 15.2 cm & 7.8 cm & 5.5 m & 2.0 cm & 0.20 & 0.59 m & 4.0 cm & 3.8 & 0.03\\
2009S & 13.4 cm & 5.4 cm & 5.2 m & 1.1 cm & 0.33 & 0.56 m & 2.2 cm & 6.2 & 0.01 \\	
2010 & 13.4 cm & 4.3 cm & 5.8 m & 0.7 cm & 0.53 & 0.61 m & 1.4 cm & 9.9 & 0.03 \\
\hline
\end{tabular}
\label{tab:1}
\end{table}

We can gain a better understanding of both the measured and the synthetic snow-depth distributions by looking at their moments. Higher order moments describe the more and more detailed structure of the probability distribution. This is precisely why we can use these higher order moments to distinguish between similar, but subtly different distributions. In Fig. \ref{fig:polashsnow}d, we compare moments of the measured snow topography and randomly-generated ``snow dune'' topographies. Higher order moments depend on the domain size, resolution, and the particular realization of the topography. For this reason, to get the moments of the simulated ``snow dune'' topographies, we created an ensemble of 50 randomly-generated topographies with the same resolution and domain size as the measurements. For each realization, we then found the root-moments, $\langle h^n \rangle^{1/n}$. Dashed lines in Fig. \ref{fig:polashsnow}d show the mean and one standard deviation of the root-moments over this 50-member ensemble. We used the model parameters shown in Table \ref{tab:1}.

For the 2010 and 2009S measurements, all of the measured moments fall squarely within the range of values obtained with an ensemble of simulated ``snow dune'' topographies. In fact, we tested the first 150 moments for these two measurements, and found that this agreement holds throughout. This means that realizations of the ``snow dune'' topography that have \textit{exactly} the same height distribution as the measurements are common. We can therefore conclude that the snow-depth distribution for 2010 and 2009S measurements is indistinguishable from the ``snow dune'' height distribution! This agreement is, however, not observed for 2009N measurements - moments of measured snow-depth distribution are consistently higher than those of the simulations and no randomly generated ``snow dune'' topography has high order moments that match the measurements. This difference is subtle, which is why we could not observe it using the KS statistic, and for most practical uses that require only a snow-depth distribution it is unlikely to be important. Nevertheless, it constitutes a real observable difference between our model and the 2009N measurements, and shows that there exist features in real data that our model cannot capture. We return to explaining this difference below.

Next, we describe the height correlation function, $C_h(l)$, for measurements and the model. We find the LiDAR $C_h(\mathbf{l})$ using Eq. \ref{autocorr}. \linelabel{A22}Due to prevailing winds, snow on sea ice often shows a strong preferential orientation, with snow dunes elongated along the direction of the wind \citep{petrich2012snow}. We observe this in the measured $C_h(\mathbf{l})$, which we find to depend on the direction of the displacement vector, $\mathbf{l}$. Such anisotropy can be included in our model by elongating the mounds along a particular axis. However, to keep our discussion as simple as possible, we chose to keep our model isotropic. Thus, to obtain a measured correlation function that depends only on the magnitude of $\mathbf{l}$, we average the measured $C_h(\mathbf{l})$ over all directions of $\mathbf{l}$. Then, we find the correlation length, $l_0$, the distance at which the measured $C_h(l)$ falls off by a factor of $e$, and relate it to the model parameter $r_0$ according to Eq. \ref{corr_length} (see Table \ref{tab:1}). We show the measured and the theoretical $C_h(l)$ calculated according to Eq. \ref{autocorr_exact} in Fig. \ref{fig:polashcorr}. 

Again, we find that 2010 and 2009S measurements agree very well with model predictions (Fig. \ref{fig:polashcorr}a). However, 2009N shows some disagreement with the model. The height correlation function calculated from raw 2009N data is shown in Fig. \ref{fig:polashcorr}b. It first decays quickly, and then the decay slows down, with features on the surface remaining correlated across the domain. By plotting $C_h$ on a semi-log plot, we find that it is approximately a sum of two exponentials - one with a short decay-length and another with a long decay-length comparable to the domain size. We get good agreement with the model after removing this long decay exponential. To do this, we fit an exponential function to $C_h$ for $l > 20 \text{m}$. We then subtract this exponential from the whole $C_h$ and rescale so that $C_h(0) = 1$. The result is a corrected correlation function, shown as the red dots in Fig. \ref{fig:polashcorr}a and the dashed line in the inset of \ref{fig:polashcorr}b. The result of this correction does not depend strongly on the choice of the cutoff length for the exponential fit so long as this length is large enough. The correlation length, $l_0$, and the model parameter $r_0$ reported in Table \ref{tab:1} correspond to this corrected correlation function.

In \citet{popovic2018simple} and SI section S1, we calculated similar correlation functions for melt pond images taken during 1998 and 2005. There, we similarly observed correlation functions with two decay lengths - one  comparable to the size of melt ponds, and another comparable to the size of the images. In \citet{popovic2018simple}, we attributed the long-decay length to differences in ice properties between different ice floes and large-scale deformation features such as ridges. A match between the model and the data could only be achieved after removing this long-decay length by a procedure analogous to the one we described above. 

Here, we again suggest that the long-decay length comes from variability in the underlying properties of the ice. In fact, we believe that all of the differences between 2009N measurements and the model can be explained by this variability in the underlying ice properties. \citet{polashenski2012mechanisms} note that the 2009N site had ice that finger-rafted early in the growth season and contained some lightly rubbled ice, while the 2009S site had very flat ice. They did not comment on ice conditions during 2010 measurements. \linelabel{Resub.338}To test whether our model is consistent with variable ice properties across the domain, we created a ``snow dune'' topography where one half of the domain had higher $h_{m,0}$ and $\rho$ than the other half, but together both parts had the same mean and variance as the measured 2009N topography. In this experiment, the model correlation function obtained a long decay length comparable to the domain size. Moreover, we found that we can choose the parameters in the first half of the domain such that all of the moments of the model height distribution match the moments of the 2009N data. Therefore, our model is consistent with the 2009N data if we assume that 2009N data contains regions with different underlying ice properties. We note that even though there exists an observable difference between the snow-depth distribution of 2009N and our model, it is very small and would likely not significantly change the ice evolution in a large scale model. Nevertheless, it hints that highly deformed ice may in fact significantly deviate from our model.

In SI section S1, we compare melt ponds on the synthetic ``snow dune'' topography with melt pond data derived from images taken during 1998 and 2005. These images cover a much larger area than the LiDAR scans we considered here. Each melt pond image covers an area of roughly 1 $\text{km}^2$ as opposed to $0.02\text{ km}^2$ covered by the LiDAR scans. As in \citet{popovic2018simple}, we find an agreement between the observed and model melt ponds. To match the melt pond statistics, we had to choose $r_0 \approx 0.6 \text{ m}$ in the model for both 1998 and 2005 ponds, in close agreement with $r_0$ we find here from the height correlation functions of the pre-melt height topographies (see Table \ref{tab:1}). In all three measurements of the LiDAR topography, we found $r_0$ within at most several cm from $r_0 = 0.6 \text{ m}$. We take this as evidence that melt ponds are controlled by the pre-melt snow topography and that the ``snow dune'' model may be meaningfully extended to at least the kilometer scale.

We will make several final notes to end this section. First, we note that since $r_0 \approx 0.6 \text{ m}$ in observations for four different years and multiple locations and ice types, the horizontal scale of the snow features seems to be a robust property of the snow cover. For this reason, future models that require a representation of snow may be able to keep the correlation length $l_0 \approx 5.6 \text{ m}$ fixed, and only keep track of mean snow depth and variance in order to represent snow in a principled way. Second, we note that distributions other than a gamma distribution that have been used to model the snow-depth, such as normal \citep{liston2018distributed} or lognormal \citep{landy2014surface}, can also often fit the measurements well. For example, a lognormal distribution passes the KS test for all three LiDAR measurements, while a normal distribution passes the KS test for 2010 measurements. However, these distributions cannot qualitatively capture the height distribution of the synthetic ``snow dune'' topography for all values of the density of mounds, $\rho$, while a gamma distribution can. For small $\rho$ (large ratio of standard deviation to mean), a normal distribution predicts a significant fraction of negative observations, while a lognormal distribution predicts a fat tail at large $h$. Since the  ``snow dune'' model so closely reproduces both the LiDAR and melt pond measurements, we believe that a distribution that is qualitatively consistent with the predictions of the ``snow dune'' model is a more justified model for a snow-depth distribution that would fit the measurements over a wider parameter range.  \linelabel{Resub.162}Finally, we note that \citet{filhol2015snow} and \citet{kochanski2019evolution} observed an approximately linear relationship between snow dune height and width for dunes that formed on snow-covered lake ice, tundra, and mountain landscapes. Our model prescription is therefore consistent with these observations. The data in \citet{kochanski2019evolution} was quite noisy, but \citet{filhol2015snow} reported a slope of the height-width relationship ranging from about 0.03 to 0.06. We can estimate the mound aspect ratio in our model as $h_{m,0}/r_0$ and compare it with \citet{filhol2015snow}. From Table \ref{tab:1}, we find that the aspect ratio of mounds in our model varies between 0.01 for 2010 measurements and 0.03 for 2009N measurements. These estimates are lower but broadly of the same order of magnitude as those of \citet{filhol2015snow}, which suggests that mounds in our model are at least broadly consistent with dunes observed in the field, although not in detail.

\section{Heat transport through the ice}\label{sec:heat}

In this section we will consider the application of our model to heat conduction through the ice. \citet{sturm2002thermal} conducted measurements on deformed multi-year ice, and concluded that the conductivity of snow inferred from large-scale ice growth is roughly 2.4 times higher than the conductivity of snow measured on-site. They ascribed a significant portion of this difference to spatial variability of snow depth and also suggested that a significant fraction of heat is transported horizontally. Motivated by their study, here we investigate the extent to which these conclusions also apply to undeformed ice. We consider a full 3-dimensional model of heat conduction to determine how much heat is extracted through undeformed ice with variable snow cover.

To model heat conduction, we assume that ice is a block of uniform thickness, $H$, that ice and snow have fixed conductivities, $k_i$ and $k_s$, that the snow cover is well-described by our ``snow dune'' model, that the temperature at the ice-ocean interface is fixed at the freezing point of salt water, $T_f$, that the temperature of the snow surface is fixed at the temperature of the atmosphere, $T_a$, and that the temperature field within the ice and snow is in a steady state. Additionally, we assume that \linelabel{Resub.398}heat is transported purely vertically in the snow (but not necessarily in the ice), which is reasonable, since snow cover is typically thin compared to its horizontal correlation length. Finally, we impose periodic boundary conditions in the horizontal directions. \linelabel{Resub.393}\linelabel{Resub.M2.2.1}Some of the assumptions above may not be realistic. For example, ice is almost certainly not a uniform block, snow conductivity, $k_s$, typically has large variations in both the horizontal and the vertical, and the system is likely not in a steady state since the conditions in the atmosphere change faster than snow and ice can adjust. For these reasons, our investigation here should be considered to be a first order estimate of the effects of snow geometry.  

In SI section S3, we develop relations for the mean conductive heat flux through the ice, $F_c$, under the assumptions above. In particular, starting from Laplace's equation for heat conduction and using the fact that our ``snow dune'' model is fully characterized by the mean, variance and the correlation length of the snow surface, we show that $F_c$ must be of the form
\begin{linenomath*}
\begin{gather}
F_c = F_0 \Phi \big( \eta, \Sigma, \Lambda \big) \text{  , }  \label{heat_flux}\\ 
F_0 \equiv k_i \frac{T_f - T_a}{H}  \text{ , }  \eta \equiv \frac{k_i}{k_s}\frac{\langle h \rangle}{H}  \text{ , } \Sigma \equiv \frac{\sigma(h)}{\langle h \rangle}  \text{ , }  \Lambda \equiv \frac{l_0}{H}  \text{  , } \label{flux_params}
\end{gather}
\end{linenomath*}
where $\Phi$ is a non-dimensional flux that is determined by a non-dimensional snow depth, $\eta$, a non-dimensional snow roughness, $\Sigma$, and a non-dimensional correlation length, $\Lambda$. The dimensional quantity $F_0$ is the heat flux conducted through the ice with no snow on top, so the function $\Phi \leq 1$ represents the fraction of this flux that is conducted when snow is present. After showing this, in SI section S3, we show that the heat conduction problem can be explicitly solved if $\Lambda \rightarrow \infty$ (when heat transport is purely vertical), and when $\Lambda \rightarrow 0$ (when the horizontal heat transport dominates). Knowing these two limits, we then numerically solve the heat conduction problem for various combinations of the parameters $\eta$, $\Sigma$, and $\Lambda$ and show that the function $\Phi$ is approximately equal to 
\begin{linenomath*}
\begin{align}
&\Phi \big( \eta, \Sigma, \Lambda \big)  \approx \Phi_v +  \frac{\Phi_h - \Phi_v}{(1+c\Lambda)^2} \text{  , }  \label{phi_tot}\\ 
& \Phi_v \equiv \int_0^\infty \frac{\tilde{f}_\Gamma(z, \Sigma)}{1+z\eta} \text{d} z  \text{ , } \label{vertical} \\ 
& \Phi_h \equiv \frac{1}{1 + \eta(1-\Sigma^2)} \text{ , } \label{horizontal} 
\end{align}
\end{linenomath*}
where $c \approx 0.83$ is a numerical constant, $\Phi_v$ is the non-dimensional flux if vertical heat transport dominates, and $\Phi_h$ is the non-dimensional flux if horizontal heat transport dominates. $\Phi_h$ is an upper bound on heat flux given parameters $\eta$ and $\Sigma$. The gamma distribution $\tilde{f}_\Gamma$ in Eq. \ref{vertical} is normalized to have a mean equal to 1, and is consequently only a function of $\Sigma$ with the parameters $k = \Sigma^{-2}$ and $h_0 = \Sigma^2$. Equation \ref{horizontal} is only valid for $\Sigma < 1$. If $\Sigma > 1$, we have that $\Phi_h = 1$ so that, if $\Lambda\rightarrow 0$, the heat is conducted as if there were no snow on top of the ice, $F = F_0$. 

To quantify how much ice growth is due to snow-depth variability, we also consider the non-dimensional flux assuming a uniform snow cover, $\Phi_u \equiv \frac{1}{1+\eta}$. The fraction of ice growth that can be attributed to snow-depth variability can then be estimated as $\frac{\Phi - \Phi_u}{\Phi}$. We show the estimated values of all of the non-dimensional parameters discussed above for the three LiDAR measurements we considered in the previous section in Table \ref{tab:2}.

From Eq. \ref{phi_tot}, we can see that the contribution of horizontal heat transport to total heat transport is proportional to $(1+c\Lambda)^{-2}$. As we discussed in the previous section, the correlation length for all datasets we considered is around $l_0 \approx 5.6\text{ m}$ and is likely on the same order throughout the Arctic. This means that for ice of realistic thickness of around $1\text{ m}$, $\Lambda \approx 5$. This implies that horizontal heat transport contributes less than 5\% of the maximum possible contribution, $\Phi_h - \Phi_v$. This is very small - as we can see from Table \ref{tab:2}, the total non-dimensional flux, $\Phi$, is equal to $\Phi_v$ to two significant digits for all measurements. Even for exceedingly thick ice of 2 m thickness, and snow with significant variability, as in the 2009N measurements, horizontal heat transport still contributes less than 1\% of total ice growth. Therefore, horizontal heat transport can most likely be ignored on flat undeformed ice. This is in contrast with \citet{sturm2002thermal} who found a significant fraction of heat is transported horizontally in regions of deformed ice. For this reason, it may be necessary to include non-uniform ice thickness to appropriately capture heat transport on deformed ice. 

Even though horizontal heat transport may be neglected on undeformed ice, snow variability may still noticeably contribute to conductive heat flux. In Table \ref{tab:2}, we show the fraction, $(\Phi - \Phi_u)/\Phi$, of heat transport that is due to snow-depth variability. We see that snow-depth variability contributes between 4\% for 2010 measurements that have the least snow-depth variability and 11\% for 2009N measurements that have the most snow-depth variability, with the effect generally increasing with $\eta$ and $\Sigma$. Although this is not large, it is comparable to the effect of melt ponds during summer - for example, if 15\% of ice of albedo 0.6 is covered by ponds of albedo 0.25, ice will melt on average 10\% faster than if there were no ponds. Therefore, the effect of snow variability on first-year ice during winter may be enough to partially compensate the effect of melt ponds during summer. Again, this effect is smaller than on deformed ice, where \citet{sturm2002thermal} found that about a quarter of ice growth is due to variable snow and ice geometry.  

\linelabel{Resub.M1.2}We note that Eqs. \ref{vertical} and \ref{horizontal}, which quantify purely vertical and purely horizontal heat flux, follow from general properties of Laplace's equation in the limits $\Lambda \rightarrow \infty$ and $\Lambda\rightarrow 0$, and from the fact that the snow depth distribution is well-described by a gamma distribution. These conditions do not change if we modify the ``snow dune'' model details such as mound shape, anisotropy, or size distribution (SI section S2.3), so our formula for total heat flux should be applicable regardless of such specifics. The only factor that could potentially change with these alterations is the detailed shape of the function that describes the transition between vertical and horizontal heat transport regimes. 

\linelabel{Resub.M2.2.2}\linelabel{Resub.393B}Finally, we note that if a reasonable statistical model of spatial variability of snow conductivity can be developed, this variability could be included in the above framework. Namely, if snow has a non-constant conductivity, $k_s(\mathbf{x})$, heat transport would be affected by an effective snow topography that accounts for these variations in conductivity, $h(\mathbf{x})\frac{\langle k_s \rangle}{k_s(\mathbf{x})}$. Thus, a non-constant snow conductivity would simply introduce an effective topography and, with a good statistical model of $k_s(\mathbf{x})$, we could follow the same steps to estimate the non-dimensional heat fluxes, $\Phi$, $\Phi_v$, and $\Phi_h$. We note that if the horizontal length-scale of snow conductivity variations is on the order of, or less than, ice thickness, it could induce significant horizontal heat transport.

\begin{table}
\setstretch{0.6}
\small
\caption{Non-dimensional parameters of the snow surface for the LiDAR measurements of \citet{polashenski2012mechanisms} assuming an ice thickness of $H=1 \text{ m}$, conductivity of fresh ice $k_i = 2.034\text{ Wm}^{-1}\text{K}^{-1}$ \citep{untersteiner1964calculations}, snow conductivity estimated by \citet{sturm2002thermal}, $k_s = 0.14\text{ Wm}^{-1}\text{K}^{-1}$, and duration of stage I of pond evolution $T = 5 \text{ days}$. }
\begin{tabular}{| c | c | c | c | c | c | c | c | c | c |}
\hline
& $\eta$ & $\Sigma$ & $\Lambda$ & $\Phi_u$ & $\Phi_v$ & $\Phi_h$ & $\Phi$ & $(\Phi - \Phi_u)/\Phi$ & $\omega$ \\
\hline		
2009N & 2.2 & 0.5 & 5.5 & 0.31 & 0.35 & 0.38 & 0.35 & 0.11 & 1.7\\
2009S & 1.9 & 0.4 & 5.24 & 0.34 & 0.37 & 0.39 & 0.37 & 0.07 & 1.7\\	
2010 & 1.9 &  0.3 & 5.75 & 0.34 & 0.36 & 0.37 & 0.36 & 0.04 & 1.5\\
\hline
\end{tabular}
\label{tab:2}
\end{table}

\section{Melt pond evolution during early melt season} \label{sec:stageI}

During the early melt season, ice is largely impermeable and ponds can quickly flood vast areas of the ice surface. This is known as stage I of pond evolution and is typically followed by stages II and III which correspond to pond drainage and subsequent slow pond growth on highly permeable ice \citep{polashenski2012mechanisms,landy2014surface}. In this section, we will discuss equations for pond evolution during stage I when ice can be considered to be impermeable, which we derive in SI section S4. These equations follow from the fact that the pre-melt snow distribution on flat ice is well-described by our ``snow dune'' topography. 

\subsection{Analytic model for pond evolution during stage I}\label{sec:drainage}

Since water can flow relatively freely through permeable snow, during stage I of pond evolution there likely exists a common water table for the entire ice floe, and ponds are the regions where surface topography lies below it. Therefore, to model stage I of pond evolution, we will assume that snow is fully permeable and that snow below the water level is completely saturated with water, that underlying ice is impermeable and initially flat, that the pre-melt snow topography is well-described by our ``snow dune'' model, and that no meltwater is lost from the domain (we will reconsider this assumption in the next subsection). The ice and snow topography can change by melting, thereby creating meltwater. We will thus assume that snow, bare ice, ponded ice, and \linelabel{A24}ponded snow (water-covered snow) melt at different rates but that these rates are constant in space and time and independent of factors such as snow or pond depth. \linelabel{B20}Finally, we will assume that both snow and ice melt only at the surface and that ice cannot melt until snow has melted away.

\begin{figure*}
\centering
\includegraphics[width=1.\linewidth]{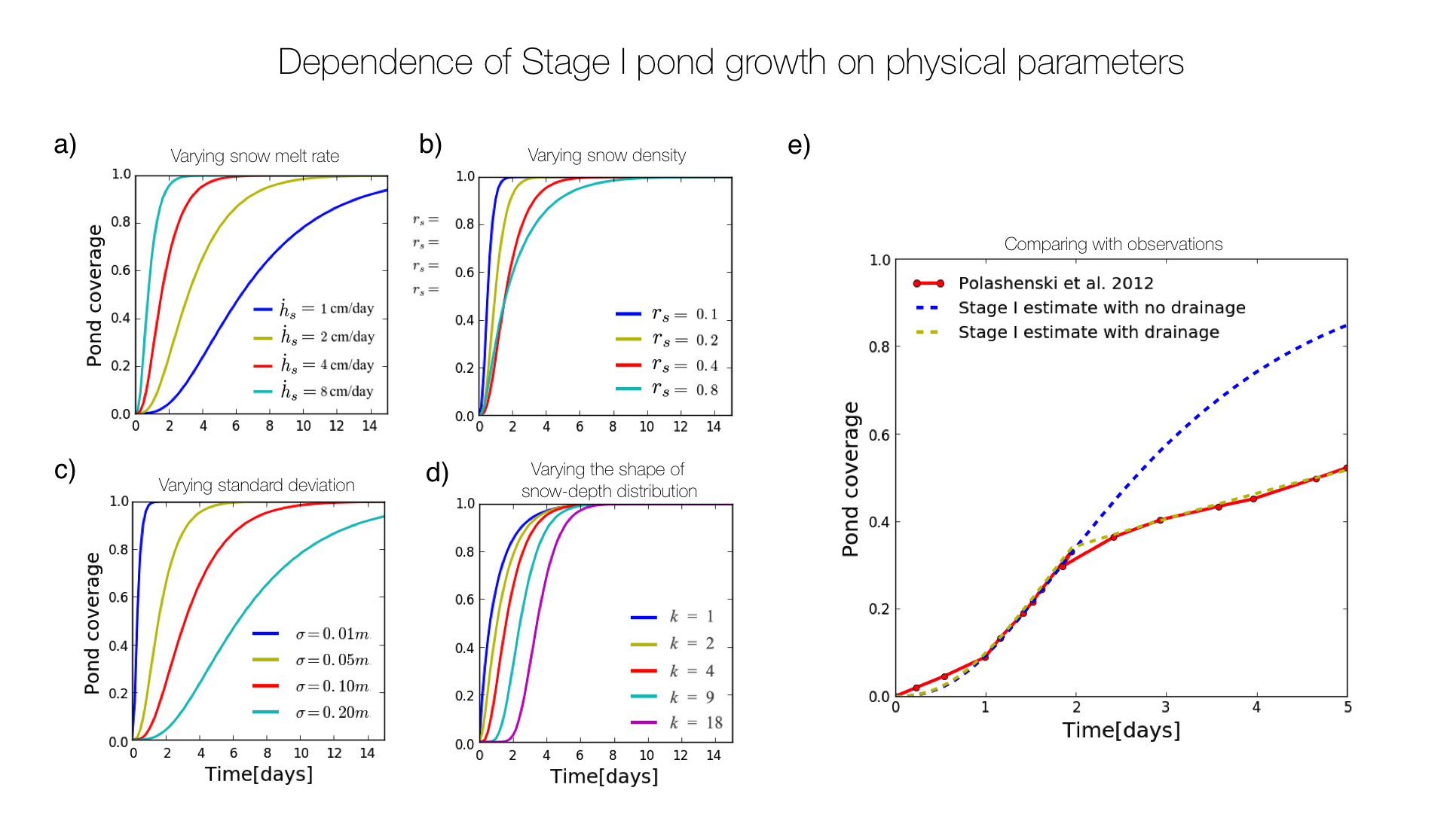}
\caption{a-d: Pond coverage evolution during stage I found using Eqs. \ref{stageIsimple} and \ref{stageIsimple_p} for a variety of model parameters. In each panel, we change one parameter, while we keep the others fixed at default values $\dot{h}_s = 4 \text{ cm day}^{-1}$, $r_s = 0.4$, $\sigma(h) = 0.05\text{ m}$, and $k = 4$.  a) Pond coverage evolution for different snow melt rates, $\dot{h}_s$. b) Pond coverage evolution for different snow to ice density ratios, $r_s$. c) Pond coverage evolution on topographies with different snow-depth standard deviations, $\sigma(h)$. d) Pond coverage evolution on topographies with different shape parameters, $k$. e) Comparing pond evolution during stage I calculated using Eqs. \ref{stageIsimple} to \ref{stageIsimple_pQ} and physical parameters described in the text (dashed blue and yellow lines) with observations of \citet{polashenski2012mechanisms} (red line). The dashed blue line represents pond evolution estimated using Eqs. \ref{stageIsimple} and \ref{stageIsimple_p} that assumes no drainage during stage I. Dashed yellow line represents pond evolution estimated using Eq. \ref{stageIsimple_pQ}, which assumes a constant drainage rate beyond the percolation threshold of $p_c = 0.35$. Topography used to calculate stage I evolution has the same mean and variance as the actual measured pre-melt snow topography obtained with LiDAR measurements.}
\label{fig:stageIparams}
\end{figure*}

In SI section S4, we develop equations for the time-evolution of pond coverage, $p$, and water level, $w$, during stage I under the assumptions above. We do this by tracking the volume of water produced by melting. We use the fact that, if no water is lost from the domain, the water level can only increase, so that ponds only encounter bare snow topography that melts at a uniform rate and is, therefore, unaltered throughout the melt season. This means that we can use our pre-melt topography to accurately describe the ponded fraction of the surface. It also means that regions of bare ice (ice with no snow and no water on top) do not exist. We greatly simplify the problem by neglecting terms that are proportional to the density difference between water and ice, which are, in addition to being small, only relevant at very high pond coverage. Using this strategy, we show that pond coverage evolution is approximately the solution to the system of ordinary differential equations
\begin{linenomath*}
\begin{eqnarray} 
&\dot{w} \approx -\frac{r_i r_s (1-p)}{1 - r_s (1 - p)}  \dot{h}_s \text{  , } \label{stageIsimple}\\
&\dot{p}  = f_\Gamma(w - \dot{h}_s t)  \Big( \dot{w} -  \dot{h}_s\Big)  \label{stageIsimple_p} \text{  , }
\end{eqnarray} 
\end{linenomath*}
where $\dot{x} \equiv \frac{\text{d} x}{\text{d} t}$ is a shorthand notation for the rate of change of quantity $x$, $w$ is the water level, $p$ is the pond coverage faction, $r_i \equiv \frac{\rho_i}{\rho_w}$ is the ratio of ice density to meltwater density, $r_s \equiv \frac{\rho_s}{\rho_i}$ is the ratio of snow density to pure ice density, $\dot{h}_s$ is the melt rate of bare snow, assumed to be constant in space and time, and $f_\Gamma(w - \dot{h}_s t)$ is the gamma distribution given by Eq. \ref{gamma_dist} and evaluated at $w - \dot{h}_s t$. The parameters of the gamma distribution are determined from the mean and variance of the pre-melt snow topography according to Eq. \ref{gammaparams}. The initial conditions for Eqs. \ref{stageIsimple} and \ref{stageIsimple_p} are no ponds at the initial time, $p(t = 0) = 0$, and a water level of zero, $w(t = 0) = 0$. Note that $\dot{h}_s < 0$, so $\dot{w} > 0$ and $\dot{p}>0$. Even though Eqs. \ref{stageIsimple} and \ref{stageIsimple_p} are approximate, they are a very accurate approximation to the full 2d model defined by the assumptions above at low pond coverage and only deviate slightly from the full model at high pond coverage (see SI section S4). In SI section S4, we also derive a second-order approximation that becomes nearly indistinguishable from the full 2d model for any pond coverage. 

We can see from Eqs. \ref{stageIsimple} and \ref{stageIsimple_p} that, since ice and water density are fixed, stage I pond coverage evolution depends mainly on the density of snow through $r_s$, the melt rate of bare snow, $\dot{h}_s$, and the mean and variance of the initial snow-depth distribution through $f_\Gamma$. Note that melt rates of bare ice, ponded ice, and ponded snow do not enter this first-order approximation for pond evolution. In Figs. \ref{fig:stageIparams}a-d, we change each of the parameters that enter Eqs. \ref{stageIsimple} and \ref{stageIsimple_p} to see how they affect stage I pond evolution. We note that in Figs. \ref{fig:stageIparams}a-d, we parameterized $f_\Gamma$ with its standard deviation, $\sigma(h)$, and its shape parameter, $k$, since $\sigma(h)$ strongly affects pond coverage evolution, while $k$ affects it only weakly. Within a reasonable range of parameters, pond evolution during stage I is most sensitive to the rate of snow melt (Fig. \ref{fig:stageIparams}a) and the standard deviation of the initial snow-depth distribution (Fig. \ref{fig:stageIparams}c). In fact, as we explain in SI section S4, increasing the snow melt rate by some factor leads to the same pond coverage evolution as decreasing the volume of snow by the same factor. Snow density and the shape of the initial snow-depth distribution, parameterized by $k$ and assuming that $\sigma(h)$ is fixed, do not matter as much (Figs. \ref{fig:stageIparams}b and d).

\subsection{Meltwater drainage during stage I}\label{sec:drainage}

\linelabel{B21}One important factor we have neglected is the potential for limited drainage during stage I. Even though ice is typically impermeable during this stage, outflow pathways such as cracks in the ice, seal breathing holes, or the floe edge can exist \citep{polashenski2012mechanisms,fetterer1998observations,eicken2002tracer,holt1985processes}. \citet{popovic2020critical} explain that if pond coverage is low, and the underlying surface is impermeable, such isolated flaws in the ice should not significantly affect the pond coverage. However, if pond coverage is above a special value called the percolation threshold, $p_c$, there can be significant pond drainage through these flaws. In fact, if the drainage rate is great enough, the pond coverage would be limited to below $p_c$. For example, unless ice deformation prevents water from flowing into the ocean, the floe edge will limit pond growth to below $p_c$. \citet{popovic2020critical} and \citet{popovic2018simple} estimate $p_c$ to be between 0.3 and 0.4 for Arctic sea ice. Below, we show that including drainage is necessary to accurately reproduce the evolution of pond coverage throughout stage I.  

\linelabel{B21.2}Including drainage in our model leads to some inconsistencies with the assumptions under which we derived Eqs. \ref{stageIsimple} and \ref{stageIsimple_p}. First, drainage opens the possibility for the water level to decrease, $\dot{w}<0$, potentially exposing bare ice and topography that was altered by different melt rates of different regions of ice. These effects were not taken into account when deriving Eqs. \ref{stageIsimple} and \ref{stageIsimple_p}, and, therefore, we can keep these equations only if the drainage is not too great. In fact, Eqs. \ref{stageIsimple} and \ref{stageIsimple_p} remain valid when $\dot{w} -  \dot{h}_s > 0$ (the altered topography remains submerged) and while $w>0$ (there are no regions of bare ice). Second, the percolation threshold only affects the pond coverage in the way we discussed above if the underlying surface is impermeable. In our case, the surface is a combination of impermeable ice and permeable snow. Since the pond coverage will likely exceed the percolation threshold before all the snow melts, there likely exists a complicated transition period between a surface that is a mix of permeable snow and impermeable ice and a fully impermeable ice surface. Moreover, when drainage is great enough so that the percolation threshold limits pond growth, the water level in different ponds needs to be at least slightly different, in contrast with the assumption of a common water table we made in the previous subsection. Here, we simply ignore these complications and assume that Eqs. \ref{stageIsimple} and \ref{stageIsimple_p} are approximately valid even when pond drainage occurs and that drainage is only activated if pond coverage exceeds the percolation threshold. We leave a detailed analysis of pond drainage during stage I for another study.

If we assume $\dot{w} -  \dot{h}_s > 0$ and $w>0$, we can straightforwardly include drainage in pond evolution (Eqs. \ref{stageIsimple} and \ref{stageIsimple_p}) by simply altering the water balance equation. We derive this in SI section S4. For example, if ponds drain at a rate of $Q$ cm per day, Eq. \ref{stageIsimple_p} remains the same and \ref{stageIsimple} is modified to 
\begin{linenomath*}
\begin{equation} \label{stageIsimple_pQ}
\dot{w} = - \frac{r_i r_s (1-p) \dot{h}_s + Q}{1-r_s(1-p)}\text{ . }
\end{equation}
\end{linenomath*}
To account for the fact that we expect little drainage to occur below the percolation threshold, we can make $Q$ a function of $p$. The simplest representation would be to include \linelabel{A26}$Q(p)= Q_0\mathcal{H}(p-p_c)$, where $\mathcal{H}$ is the Heaviside function, equal to 0 or 1 depending on whether its argument is less or greater than 0, and $Q_0$ is a constant that sets the maximum drainage rate. Using this function yields the yellow line in Fig. \ref{fig:stageIparams}e. If $\dot{w} -  \dot{h}_s < 0$, Eq. \ref{stageIsimple_p} predicts that the pond coverage will decrease, $\dot{p} < 0$. So, if $\dot{w}  <  \dot{h}_s$, we expect the pond coverage to remain fixed at the percolation threshold, $p = p_c$.

In Fig. \ref{fig:stageIparams}e we compare stage I pond evolution predicted by our model with the measurements of stage I pond coverage evolution by \citet{polashenski2012mechanisms}. Since accurate measurements of pre-melt topography are available for that time and location (Fig. \ref{fig:polashsnow}), we can choose the parameters of the gamma distribution highly accurately. The density of snow and the snow melt rate are uncertain. We choose the density of snow to be $\rho_s = 350 \text{ kg m}^{-3}$, and the albedo of melting snow to be $\alpha_s = 0.74$, consistent with observations of \citet{polashenski2012mechanisms}. We take the solar flux to be $F_\text{sol} = 254 \text{ W m}^{-2}$, which \citet{polashenski2012mechanisms} report as the average flux of solar energy for the duration of the experiment. We treat the sum of longwave, sensible, and latent heat fluxes, $F_r$, as a tuning parameter. The value we choose, $F_r  = -15 \text{ W m}^{-2}$, is typical of the region and season, and is therefore consistent with the measurements. The blue line in Fig. \ref{fig:stageIparams}e shows the stage I pond evolution predicted by Eqs. \ref{stageIsimple} and \ref{stageIsimple_p}, while the red line represents the measurements of \citet{polashenski2012mechanisms}. The two curves agree up to the point when pond coverage reaches $p \approx 0.35$, a plausible value for the percolation threshold. Thus, the measurements are consistent with drainage being activated beyond the percolation threshold. \linelabel{A26.2}To match pond evolution beyond $p = 0.35$ we used Eq. \ref{stageIsimple_pQ}, and tuned the drainage rate, $Q_0$, to $2.6\text{ cm day}^{-1}$ (yellow line), \linelabel{B22}a rate consistent with observations of \citet{polashenski2012mechanisms}. We can see that Eq. \ref{stageIsimple_pQ} is consistent with the observations, although this matching cannot be fully confirmed due to uncertainty in the parameters.   

\subsection{A condition for pond development}\label{sec:effects}

\begin{figure*}
\centering
\includegraphics[width=0.6\linewidth]{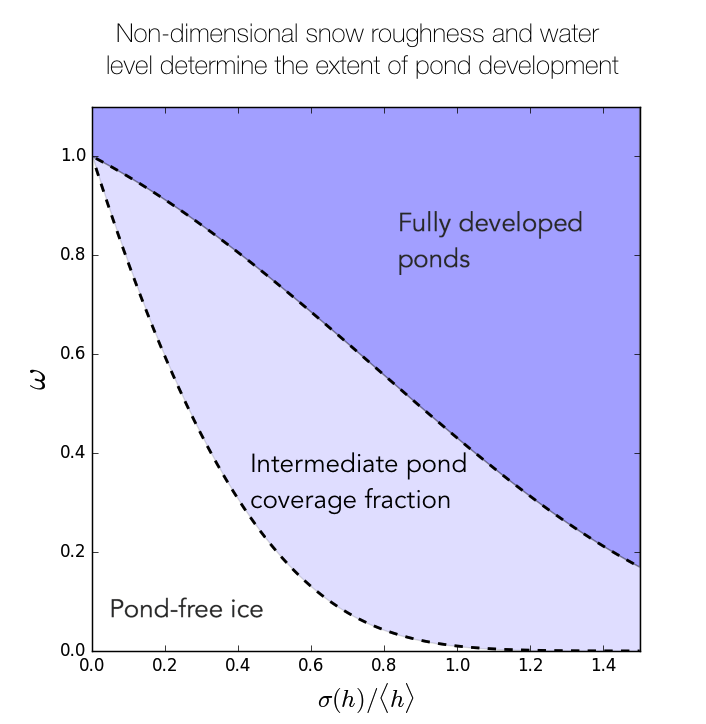}
\caption{The horizontal axis shows the non-dimensional snow roughness, $\sigma(h)/\langle h \rangle$, while the vertical axis shows the non-dimensional water level, $\omega$, described in the text. The white-colored region represents ice that remains pond-free throughout the summer, while the dark-blue region represents ice that develops pond coverage that exceeds the percolation threshold by the end of stage I. The light-blue region represents ice that has some ponds by the end of stage I, but the pond coverage does not exceed the percolation threshold. }
\label{fig:phase_diagram}
\end{figure*}

Pond coverage reaches its peak by the end of stage I. Afterwards, ponds drain, and, by the end of the drainage stage, the remaining ponds correspond to regions of ice that are below sea level. Observations show that ponds that remain after drainage are a subset of ponds that exist during peak pond coverage \citep{polashenski2012mechanisms,landy2014surface}. Therefore, if ponds do not develop during stage I, they will likely never develop. Since the pre-melt snow-depth distribution controls pond evolution during stage I, it, therefore, also has a disproportionate effect on later pond evolution. If the snow-depth is highly variable, ponds will develop sooner than if the snow cover is uniform. If the snow cover is relatively uniform and stage I does not last long enough, the ice may remain pond-free throughout the summer. Here, we will show how we can use the model we developed above to derive a simple criterion for whether ice will develop ponds or not. 

We start with Eqs. \ref{stageIsimple} and \ref{stageIsimple_p}. If we assume that pond coverage stays low throughout stage I, we can set $p \approx 0$ in Eq. \ref{stageIsimple}, and integrate it directly to get the water level at time $t$ during stage I, $w(t) \approx -\frac{r_i r_s }{1 - r_s}  \dot{h}_s t$. We can then non-dimensionalize Eq. \ref{stageIsimple_p} by introducing non-dimensional parameters $\Sigma$ and $\omega$ to get
\begin{linenomath*}
\begin{align} 
&\frac{\text{d} p}{\text{d} \omega} = \tilde{f}_\Gamma(\omega,\Sigma)  \text{  , } \label{p_eta} \\
\Sigma \equiv \frac{\sigma(h)}{\langle h \rangle} \quad \text{ , } &\quad \omega \equiv - \frac{1- r_s(1-r_i) }{1 - r_s} \frac{\dot{h}_s t}{\langle h \rangle} \text{  , }
\end{align} 
\end{linenomath*}
Here, $\Sigma$ is the non-dimensional roughness as in section \ref{sec:heat} and $\omega$ can be interpreted as a non-dimensional water level. As in Eq. \ref{vertical} of section \ref{sec:heat}, $\tilde{f}_\Gamma(\omega,\Sigma)$, is a gamma distribution with a mean equal to 1, so that it only depends on $\Sigma$, with parameters $k=\Sigma^{-2}$ and $h_0 = \Sigma^2$, and evaluated at $\omega$. Thus, from Eq. \ref{p_eta}, we get a simple relation 
\begin{linenomath*}
\begin{equation} 
p = F_\Gamma(\omega,\Sigma)  \text{  , } \label{cumulative_p_eta}
\end{equation} 
\end{linenomath*}
where $F_\Gamma(\omega,\Sigma)$ is a cumulative gamma distribution with a mean equal to 1, $F_\Gamma( \omega,\Sigma) \equiv \int_0^{\omega} \tilde{f}_\Gamma(z,\Sigma)\text{d} z$. Therefore, if pond coverage is low, it depends only on $\omega$ and $\Sigma$ and we can express it simply in terms of the cumulative gamma distribution. 

Let us now assume that stage I lasts for some time, $T$, on the order of 5 days. If the ponds do not develop within that time, the ice will remain pond-free throughout the summer. We can set some threshold, $p^*$, say $p^* = 0.01$, and consider ice to be pond-free if the pond coverage is below $p^*$. We can then find the boundary in $\omega$-$\Sigma$ space, $\omega^*(\Sigma)$, such that if $\omega < \omega^*$, ice remains pond-free at the end of stage I. According to Eq. \ref{cumulative_p_eta}, this boundary is given by
\begin{linenomath*}
\begin{equation} 
\omega^*(\Sigma) = F^{-1}_\Gamma(p^*,\Sigma)  \text{  , } \label{criterion}
\end{equation} 
\end{linenomath*}
where $F^{-1}_\Gamma$ is an inverse of the cumulative gamma distribution (gamma percentile function) with a mean equal to 1, $k=\Sigma^{-2}$, and $h_0 = \Sigma^{2}$, evaluated at $p^*$. Equation \ref{criterion} gives a universal criterion that ice remains pond-free. If $\omega$, calculated given the snow melt rate, density, depth, roughness, and the duration of stage I, exceeds $\omega^*$, ice will develop at least some ponds. The boundary, $\omega^*(\Sigma)$, is largely insensitive to the choice of the threshold coverage, $p^*$, so long as $p^*$ is much smaller than 1. 

In the case when ice develops some ponds, we can also ask how developed those ponds will be. In particular, we can assume that ponds are fully developed if the pond coverage exceeds the percolation threshold by the end of stage I. Beyond the percolation threshold, pond drainage likely starts to play an important role, and the post-stage I evolution of fully-developed ponds likely proceeds in a fairly typical manner. Therefore, there likely exist three typical trajectories for pond evolution during summer - pond-free, fully-developed, and intermediate. To see whether the ponds will develop fully, we can again use Eq. \ref{criterion} but use $p^* = p_c \approx 0.35$. Using Eq. \ref{criterion} with $p^* = 0.35$ is approximate since the pond coverage is no longer small, but it is still reasonably accurate. 

In Fig. \ref{fig:phase_diagram}, we show the conditions for these three regimes of pond development. We can see that when $\omega > 1$, ponds always fully develop since, in this case, the water level by the end of stage I exceeds the height of snow even when the snow cover is uniform. Otherwise, if $\omega < 1$, ponds develop more easily when the snow-depth has more variability. We can estimate $\omega$ and $\Sigma$ for the measurements of \citet{polashenski2012mechanisms}. We use $T = 5\text{ days}$, snow depth mean and variance estimated from the LiDAR measurements, and the same snow melt rate density as in the previous subsection. We show $\Sigma$ and $\omega$ estimated in this way in Table \ref{tab:2}. We can see that $\omega > 1$ in all cases so that we expect that ponds develop fully. In the study of \citet{polashenski2012mechanisms} this indeed happened. In order for ice to have remained pond-free, stage I would have had to have lasted less than 1.5 days in 2010, which had the most uniform snow, and less than 0.5 days for 2009N, which had the most variable snow. We note that persistently pond-free ice has been observed in the Arctic \citep{perovich2002aerial}, while Antarctic sea ice only rarely develops ponds. \linelabel{Resub.655}Although we do not have snow measurements from these pond-free ice floes, general conditions on Antarctic ice are broadly consistent with our framework here. Snow on Antarctic ice is typically deeper than in the Arctic, lowering both $\Sigma$ and $\omega$, and it is both thinner and more saline than Arctic ice, potentially lowering the duration of stage I, $T$, thereby further lowering $\omega$. It would be interesting to see where these pond-free ice floes fall within our pond-development diagram.

\section{Discussion}\label{sec:discussion}

\linelabel{Resub.656}\linelabel{Resub.M.3}Throughout this paper, we showed that our model can reproduce key statistical features on three nearby locations where detailed LiDAR data were collected as well as two other locations where melt pond images were taken. \linelabel{Resub.665}Although this is not enough to claim generality, our results are encouraging, suggesting that our model may be applicable in large-scale studies of snow on undeformed Arctic sea ice and could serve as a starting point for studying snow on other terrains such as deformed ice, Antarctic ice, or Alpine snow fields. \linelabel{Resub.667}Moreover, as the fraction of first-year ice increases in the Arctic due to global warming, the conditions similar to the ones we tested here will likely become more common. Here, we will briefly discuss how our results can be applied to large-scale studies, assuming their generality holds.

\begin{enumerate}

\item We believe that large-scale models should use a gamma distribution for snow depth on undeformed ice. Even though other distributions, such as normal or lognormal, may still work well enough, adopting a gamma distribution is a more principled approach that likely works over a wider range of parameters. Therefore mean and variance of snow depth are enough to characterize the whole distribution according to Eqs. \ref{gammaparams}. \linelabel{Resub.M.4}Since, according to the LiDAR as well as melt pond data, the correlation length appears to be stable under a variety of environmental conditions, it may be possible to use our ``snow dune'' model with a horizontal scale parameter $r_0 = 0.6\text{ m}$ ($l_0 = 5.6\text{ m}$) to study horizontal snow distribution. 

\item When modeling heat conduction through undeformed ice, it is likely safe to neglect horizontal heat transport. On deformed ice, the situation may be more complicated and further study is required. 

\item To parameterize ice albedo during summer, Eqs. \ref{stageIsimple} and \ref{stageIsimple_p} may be used for pond evolution during stage I. Without knowing the outflow rate, $Q$, a reasonable approach would be to cap pond coverage to below the percolation threshold of 0.3 to 0.4. Equations \ref{stageIsimple} and \ref{stageIsimple_p} can be combined with similar approaches, such as the model of \citet{popovic2020critical} for stages II and III, or the model of \citet{popovic2018simple} for stage III, to yield an analytic and accurate model of pond evolution throughout the entire melt season.

\item \linelabel{Resub.676}Eq. \ref{criterion} provides a yet-untested prediction of whether ice will remain pond-free throughout the summer - we were not able to test this criterion against real data appropriately, as in all three cases we considered, the ponds have fully developed. So, it would be interesting to see whether this criterion can explain the observations of pond-free ice in the Arctic and the Antarctic. The non-dimensional water level, $\omega$, in Eq. \ref{criterion} depends on the duration of stage I. So, to be able to use this equation in large scale-studies, it will be necessary to relate the duration of stage I to parameters that are available on the large scale, such as energy fluxes, thickness, or the salinity of ice.

\end{enumerate}

\section{Conclusions}\label{sec:conclusions}

Snow cover greatly impacts sea ice evolution. It insulates the ice during the winter, reflects sunlight during the summer, and controls melt pond evolution. \linelabel{Resub.685}In this paper, we presented an idealized geometric model of ``snow dune'' topography that was capable of capturing key statistical properties of the snow cover on undeformed first-year sea ice on three sites for which detailed LiDAR measurements were available as well as for two other missions that collected airborne melt pond images. The main conclusion of our paper is that, at least for the measurements we considered, our ``snow dune'' model is a very accurate representation of the 3-dimensional snow distribution on flat, undeformed Arctic sea ice.

By comparing the moments of the height and snow-depth distributions, we showed that the height distribution of our model is statistically indistinguishable from the snow-depth distribution measured in detailed LiDAR scans on flat undeformed ice. We also compared the correlation functions for the model and the measurements to show that the horizontal statistics of snow are also well-captured by our model. In addition to this, our model captures melt pond geometry derived from helicopter images that span areas of 1 km$^2$ with the same model parameters as the LiDAR scans. This suggests that our model may capture the horizontal properties of snow on such large scales. We note that in all cases we tested, the correlation length of horizontal features (either snow or melt pond), was around $l_0\approx 5.6\text{ m}$, suggesting that this property \linelabel{Resub.701}may be constrained throughout the Arctic. Since our model is fully defined by only three parameters, it follows that it is enough to know the mean snow depth, its variance, and the horizontal correlation length to fully characterize the statistics of snow on undeformed ice. On ice that was slightly deformed we showed that the measured snow-depth distribution contains subtle differences from the model and that there exist long-range correlations in the horizontal, inconsistent with our model.

After comparing our ``snow dune'' model to measurements, we considered its application to heat conduction through the ice and to the development of melt ponds. We solved a 3-dimensional heat conduction equation assuming the ice is flat and snow cover that is well-described by our model. We developed simple formulas that solve the conductive heat flux problem for any configuration of snow consistent with our model. Using these formulas, we showed that, \linelabel{Resub.713}for any realistic ice thickness and snow dune size, horizontal heat transport may be neglected on flat, undeformed ice (although this may not be the case for deformed ice). We then also used our model to develop a simple model for melt pond evolution during early stages of pond development when ice can be considered impermeable. We find that pond coverage evolution can be estimated using a system of two coupled ordinary differential equations that closely approximate an equivalent 2d model, but are much simpler to solve. They reveal connections with measurable environmental parameters. For example, we show that doubling the amount of solar radiation has approximately the same effect on pond growth rate as halving the snow depth. We also show that they are consistent with observations. Using these equations, we develop a criterion in terms of non-dimensional properties of the ice surface for whether ice will develop ponds or remain pond-free during summer.

In this paper, we demonstrated that \linelabel{Resub.725}many of the details of ice surface conditions can be summarized by only a few parameters if one is interested in bulk properties such as the total heat conducted or the mean melt pond coverage. \linelabel{Resub.728}Of the three parameters that define our model, it may be the case that only two change from situation to situation, as the correlation length may be stable, as discussed above. This means that future GCMs may need to only keep track of the total amount of snow and the snow depth variance in order to be able to represent the snow surface in a principled way. Our results, therefore, uncover an important property of sea ice snow cover that may improve the realism of sea ice models \linelabel{Resub.733}while adding little to their complexity.

\acknowledgments
We thank BB Cael and Stefano Allesina for reading the paper and giving comments. We thank Don Perovich for providing the melt pond image in Fig. \ref{fig:topos}d. Predrag Popovi\'{c} was supported by a NASA Earth and Space Science Fellowship. This work was partially supported by the National Science Foundation under NSF award number 1623064 and by National Science Foundation award DMS-1517416 (MS). Code to generate the synthetic ``snow dune'' topography is archived as \citet{pedjapopovic20203930517} and is also available at \url{https://github.com/PedjaPopovic/Snow-dune-topography}. LiDAR measurements collected by \citet{polashenski2012mechanisms} are available for download at \url{http://chrispolashenski.com/data.php} or \url{https://data.eol.ucar.edu/dataset/106.426}. The authors declare no conflict of interest.

\bibliography{mybibliogJGR}

\end{document}


\pagestyle{plain}

\title{Supplementary Information for ``Snow topography on undeformed Arctic sea ice captured by an idealized `snow dune' model''}

\authors{Predrag Popovi\'{c}\affil{1}, Justin Finkel\affil{2}, Mary C. Silber\affil{2}, Dorian S. Abbot\affil{1}}

\affiliation{1}{The University of Chicago, The Department of the Geophysical Sciences}
\affiliation{2}{The University of Chicago, Department of Statistics and Committee on Computational and Applied Mathematics}

\correspondingauthor{Predrag Popovi\'{c}}{arpedjo@gmail.com}


\renewcommand{\theequation}{S\arabic{equation}}
\renewcommand{\thefigure}{S\arabic{figure}}
\renewcommand{\thesection}{S\arabic{section}}
\renewcommand{\thesubsection}{S.\arabic{section}.\arabic{subsection}}
\renewcommand{\thesubsubsection}{S.\arabic{section}.\arabic{subsection}.\arabic{subsubsection}}

\section{Melt pond geometry}\label{sec:geom}

\begin{figure}
\centering
\includegraphics[width=0.8\linewidth]{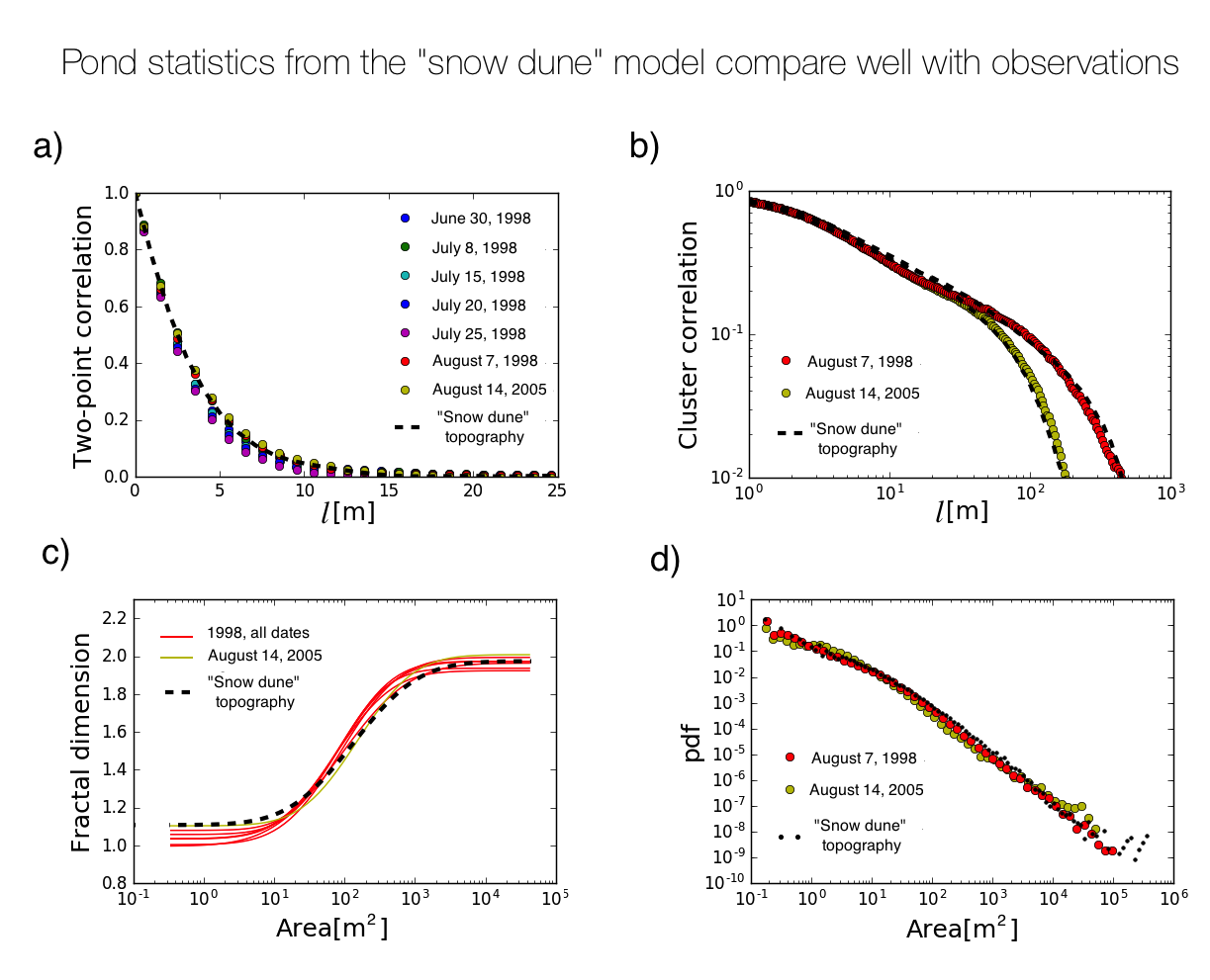}
\caption{A comparison of geometric properties of ponds photographed on different dates during the 1998 SHEBA and the 2005 HOTRAX missions with the geometry of ponds on the synthetic ``snow dune'' topography. The synthetic ``snow dune'' topography has a mean horizontal scale of mounds of $r_0 = 0.6 \text{ m}$ and the pond coverage fraction is set close to the percolation threshold. a) Two-point correlation function. Dots represent measurements, while the black dashed line represents synthetic topography. b) Cluster correlation function. Dots represent measurements, while the black dashed line represents synthetic topography. The two dashed black lines differ only in the domain size, with the domain sizes corresponding to the image sizes of the SHEBA and HOTRAX images. c) Fractal dimension of pond boundary as a function of pond area. Red and yellow lines are measurements, while the black dashed line represents ponds on synthetic topography. d) Pond area distribution on a log-log plot. Red and yellow dots are measurements, while the black dots represent ponds on synthetic topography   }
\label{fig:geom}
\end{figure}

In this section we confirm that the melt pond features on the synthetic ``snow dune'' topography reproduce Arctic sea ice melt pond statistics with as much accuracy as the ``void model'' of \citet{popovic2018simple}. We do this by comparing melt ponds on the synthetic ``snow dune'' topography to ponds from the same dataset used by \citet{popovic2018simple}. Figure \ref{fig:geom} shows that the comparison between the model and the data is nearly identical to that of \citet{popovic2018simple} - the only difference is that the model curves now represent the synthetic ``snow dune'' topography instead of the ``void model.''

In \citet{popovic2018simple}, we compared the statistics of the ``void model'' to the statistics of real ponds derived from photographs taken during the 1998 SHEBA mission and the 2005 HOTRAX missions. We compared four statistical features between the model and the data: the two-point correlation function, $C(l)$, the cluster correlation function, $G(l)$, the fractal dimension of the pond boundary, $D(A)$, as a function of pond area, $A$, and the pond area distribution, $f(A)$. Here we will use the same dataset and the same statistics to validate the synthetic ``snow dune'' topography. Below we explain each of these statistics and the method we used to obtain them only briefly, and we point the reader to \citet{popovic2018simple} for details.

We form ponds on the synthetic ``snow dune'' topography by intersecting the surface with a flat plane (a ``water table'') and assuming ponds are regions of the surface below this plane. In this way, the pond coverage fraction can be changed independently of the surface parameters by simply raising and lowering the water table. This is in contrast with the ``void model,'' where the density of circles placed controls the fraction of the surface covered by ponds, somewhat complicating the analogy between the two models. This will, however, not be a problem as pond statistics turn out to depend only on the typical radius, $r_0$, and the deviation of pond coverage from the percolation threshold, $p_c$, in both the ``void model'' and the ``snow dune'' model. We explain the meaning of the percolation threshold below. Therefore, to compare the ``snow dune'' model with real ponds, we will use two of the statistics ($C(l)$ and $G(l)$) to calibrate $r_0$ and $p-p_c$ in the model, and use the other two statistics to confirm that the model reproduces the data with the same parameters. 

A two-point correlation function, $C(l)$, is defined via the probability, $P(l)$, that, given a randomly chosen point on a pond, a point distance $l$ away is also located on a pond. Precisely, the two point correlation function is given as 
\begin{linenomath*}
\begin{equation}
C(l) = \frac{P(l) - p}{1-p} \text{ . }
\end{equation}
\end{linenomath*}
This function is mostly sensitive to the horizontal scale of the mounds, $r_0$, and can be used to calibrate this parameter. The two-point correlation function for pond photographs is approximately a sum of two exponentials, similar to the height correlation function we discussed in section IV of the main text. The exponential with a short decay scale most likely corresponds to melt ponds, while the exponential with a long decay scale likely corresponds to large-scale features such as ridges and floe edges. To be able to compare the model to data, we have to remove the long-decay exponential in the same way as we did in section IV of the main text (see also \citet{popovic2018simple} for details). After removing the long-decay exponential, and choosing $r_0 = 0.6\text{ m}$, we get good agreement for $C(l)$ between the model and the data (Fig. \ref{fig:geom}a). Note that $r_0 = 0.6\text{ m}$ is very close to the $r_0$ we recovered from the LiDAR scans in section IV of the main text (see Table I of the main text).

A cluster correlation function, $G(l)$, is defined as the probability that, given a randomly chosen point on a non-spanning pond, a randomly chosen point a distance $l$ away from this point is located on the \textit{same} pond. A non-spanning pond is a pond that does not span the entire domain. This function is sensitive to the size of the largest non-spanning clusters, and therefore, to the pond coverage fraction, $p$. As the pond coverage fraction is increased, the largest connected cluster grows. At first it does not span the domain but after the pond coverage exceeds a special value called the percolation threshold, $p_c$, the largest cluster begins spanning the domain and therefore does not enter into the calculation of $G(l)$ anymore. It turns out that $G(l)$ depends on the deviation of pond coverage from the percolation threshold, $p - p_c$. For this reason, it can be used to calibrate $p - p_c$. In the void model, the percolation threshold is fixed, $p_c \approx 0.33$, while in the ``snow dune'' model, $p_c$ depends on the density of mounds, $\rho$. So, we fix $\rho = 0.2$ and then we find the $p$ for which $G(l)$ for the model matches $G(l)$ for the data. Finding the pond coverage at which spanning ponds first appear, we find that the percolation threshold for $\rho = 0.2$ is approximately $p_c \approx 0.44$. We find that for $p-p_c \approx 0.015$, $G(l)$ for the model and data agree well (Fig. \ref{fig:geom}b). So, the pond coverage that reproduces $G(l)$ for the data is close to, but slightly above the percolation threshold. The fact that ponds seem to be organized near the percolation threshold is in agreement with the observations of \citet{popovic2018simple} and the theoretical considerations of \citet{popovic2020critical}. 

Having calibrated $r_0$ and $p-p_c$, we proceed to test whether $D(A)$ and $f(A)$ agree between the model and data using the same parameters of the surface. Following \citet{hohenegger2012transition} and using a method described in \citet{popovic2018simple}, we determined the fractal dimension of the pond boundary as the exponent, $D$, that relates the pond area, $A$, to its perimeter, $P$, $P \propto A^{D/2}$. We treat $D$ as a function of pond area. In Fig. \ref{fig:geom}c, we show that the fractal dimension for synthetic ponds reproduces the fractal dimension of real ponds. Finally, we look at the pond area distribution, $f(A)$. Again, using the parameters chosen by calibrating the model with the correlation functions, we find that the area distribution for real and synthetic ponds agree over the entire observational range (Fig. \ref{fig:geom}d).

We also test the model for different values of $\rho$ and varying the anisotropy of mounds. We find that, as long as $p - p_c$ is the same, agreement between the data and the model remains. We take the accurate agreement between the ``snow dune'' model and the melt pond data along with the fact that the parameter $r_0$ is very similar for melt pond images and the LiDAR scans of the snow surface to be evidence that the horizontal characteristics of snow are well-represented in our model even on large scales. Melt pond photographs are taken at the end of summer, whereas ponds forming with a common water table on a snow-covered surface would be consistent with ponds during early summer. However, since ponded ice melts faster than bare ice and snow melts slower than bare ice, it is likely that the initial snow topography is approximately preserved throughout the melt season. Nevertheless, this difference in timing likely contributes to the fact that we estimated the percolation threshold for real ponds to be between 0.3 and 0.4 \citep{popovic2018simple}, while the ``snow dune'' topography predicts $p_c$ between 0.4 and 0.5 for different values of $\rho$. A lower percolation threshold for real ponds could be due to the fact that ponds form channels as they drain during the later stages \citep{polashenski2012mechanisms,landy2014surface}, thereby increasing the connectivity and lowering $p_c$.

\section{Detailed statistics of the synthetic ``snow dune'' topography}\label{sec:moments}

In this section, we first prove Eqs. 4, 5, and 7 of the main text that we showed in section III of the main text. Then we discuss the relation between the height distribution of the ``snow dune'' topography and the gamma distribution. Finally, we test the sensitivity to model statistics to details such as the shape of the mounds and the distribution of mound radii. 

\subsection{Proving the results of section III of the main text}

\subsubsection{Proof of Eqs. 4 and 5 of the main text}

Here we derive the recursion relation, Eq. 5 of the main text, for the moments of the synthetic ``snow dune'' topography. Equations 4 of the main text for the mean and the variance are then a corollary of this result. This proof has two steps. The first is to show where the recursion relation comes from, and the second is to find the moments of individual mounds. The first step will turn out to be universal, while the second will contain elements that depend on the details of the model. Throughout the derivation, we will use the notation $\langle \text{ ... } \rangle$ for the mean of some function of topography. This mean can equally be interpreted as the mean over the domain or, since we are assuming that the domain is infinite, as the mean with respect to the probability distributions of radii, $r_i$, and locations, $\mathbf{x}_{0,i}$, of individual mounds. In this section, we will only take ensemble averages to derive the theoretical moments of the ``snow dune'' height distribution, whereas in the main text we took spatial averages to empirically find the moments of computer-generated synthetic topographies.

First, we show where the recursion relation, Eq. 5 of the main text, comes from. The ``snow dune'' topography is a sum of individual mounds
\begin{linenomath*}
\begin{equation}\label{toposum}
h_\text{SD}(\mathbf{x}) = \sum_{i=1}^{N} h_i(\mathbf{x}) \text{  , }
\end{equation}
\end{linenomath*}
where $h_i(\mathbf{x})$ is the height of the $i$-th mound $h_i(\mathbf{x}) \equiv  h_{m,0} \frac{r_i}{r_0} e^{-(\mathbf{x}-\mathbf{x}_{0,i})^2 / 2 r_i^2}$. Therefore, the $n$-th power of the topography is 
\begin{linenomath*}
\begin{equation}\label{toposum}
h_\text{SD}^n(\mathbf{x}) = \sum_{i_1 \text{...} i_n = 1}^{N} h_{i_1}(\mathbf{x})\text{ ... } h_{i_n}(\mathbf{x}) \text{  , }
\end{equation}
\end{linenomath*}
The $n$-th moment of the topography can then be found as $\langle h_\text{SD}^n \rangle$. However, before we can find the mean of the above sum, we need to rearrange it into a different form. Since the mounds are sized and placed independently, we can use the identity $\langle h_i(\mathbf{x}) h_j(\mathbf{x})\rangle = \langle h_i(\mathbf{x})\rangle \langle h_j(\mathbf{x})\rangle $ for $i \neq j$, with $\langle \text{ ... } \rangle$ being the ensemble average over all possible locations and radii of mounds $ h_i$ and $ h_j$. So, in the sum above (Eq. \ref{toposum}), we need to distinguish between the terms where $i = j$ and the terms where $i \neq j$. To this end, we can first split the sum into a term where all $n$ indexes are equal and a term where at least one of the $n$ indexes is different from the others. We thus get 
\begin{linenomath*}
\begin{equation}\label{firstsplit}
\sum_{i_1 \text{...} i_n}^{N} h_{i_1}\text{ ... } h_{i_n} = \sum_{i=1}^{N} h_{i}^n +  \sideset{}{'}\sum_{i\text{;} i_1 \text{...} i_{n-1} }^{N} h_i h_{i_1}\text{ ... } h_{i_{n-1}}  \text{  , }
\end{equation}
\end{linenomath*}
where the prime on the second sum on the right hand side means that at least one of $i_1 \text{ ... } i_{n-1}$ is not equal to the first index, $i$. It is now straightforward to take the average of the first term where all the terms are equal, but the average of the second term is still problematic. Since the second sum has at least one index different from from the first index $i$, we can further split it into a sum that has exactly one index different from $i$ and a sum where there are at least two indexes different from $i$ 
\begin{linenomath*}
\begin{equation}\label{secondsplit}
\sideset{}{'}\sum_{i\text{,} i_1 \text{...} i_{n-1} }^{N} h_i h_{i_1}\text{ ... } h_{i_{n-1}} = (n-1) \sum_{i \text{;} j\neq i}^{N} h_{i}^{n-1} h_j + \sideset{}{''}\sum_{i\text{;} i_1 \text{...} i_{n-1} }^{N} h_i h_{i_1}\text{ ... } h_{i_{n-1}}    \text{  , }
\end{equation}
\end{linenomath*}
where now $ \sideset{}{''}\sum$ means that there are at least two terms among $i_1 \text{ ... } i_{n-1}$ not equal to $i$. The first sum above has exactly one term, $j$, not equal to $i$, and the factor $(n-1)$ comes from the fact that $j$ can correspond to any of the $i_1 \text{ ... }i_{n-1}$ terms in the sum $\sideset{}{'}\sum h_i h_{i_1}\text{ ... } h_{i_{n-1}}$. The idea is then to continue splitting the primed sum in this way until we end up with a sum where all the indexes are different. We do this as follows. After $s$ steps, we end up with a primed sum that has at least $s$ terms among $i_1 \text{ ... } i_{n-1}$ not equal to $i$. We then split this sum to get a sum where exactly $s$ terms are not equal to $i$ and a sum where at least $s+1$ terms are not equal to $i$
\begin{linenomath*}
\begin{equation}\label{ssteps}
\sideset{}{^{(s)}}\sum_{i\text{,} i_1 \text{...} i_{n-1} }^{N} h_i h_{i_1}\text{ ... } h_{i_{n-1}} = {n-1 \choose s} \sum_{i , j_1\neq i,\text{...}j_s\neq i}^{N} h_{i}^{n-s} h_{j_1}\text{...}h_{j_s} + \sideset{}{^{(s+1)}}\sum_{i\text{,} i_1 \text{...} i_{n-1} }^{N} h_i h_{i_1}\text{ ... } h_{i_{n-1}}    \text{  . }
\end{equation}
\end{linenomath*}
Therefore, we end up with one term where none of $ j_1\text{ ... }j_s$ are equal to $i$ (note that there is no condition on whether or not the indexes $j_1\text{ ... }j_s$ are equal to each other). The factor ${n-1 \choose s}$ comes from the number of ways that we can choose $s$ indexes $ j_1\text{ ... }j_s$ from the set $ i_1\text{ ... } i_{n-1}$. Finally, after $n-1$ steps, we arrive at the end of this process. The final term is thus a sum where none of $ j_1\text{ ... }j_{n-1}$ are equal to the first term, $i$. At this step, we can proceed to take the average, $\langle h_\text{SD}^n \rangle$. Using the fact that in each sum, the first term, $i$, is different than all the other terms, we get
\begin{linenomath*}
\begin{equation}\label{mthmoment}
\begin{split}
\big\langle h_\text{SD}^n \big\rangle = \sum_{i=1}^{N} \big\langle h_{i}^n \big\rangle + \text{...} +&{n-1 \choose s} \sum_{i=1}^{N} \big\langle h_{i}^{n-s} \big\rangle  \Big\langle \sum_{j_1\text{...}j_s\neq i}^{N} h_{j_1}\text{...}h_{j_s} \Big\rangle + \\
& \text{...} + \sum_{i=1}^{N} \big\langle h_{i} \big\rangle \Big\langle \sum_{j_1\text{...}j_{n-1}\neq i}^{N} h_{j_1}\text{...}h_{j_{n-1}} \Big\rangle \text{  . }
\end{split}
\end{equation}
\end{linenomath*}
First, we can note here that terms $\langle h_{i}^{n-s} \rangle$ are equal for all $i$, since all mounds are drawn from the same distribution. So, the terms $\sum_{i=1}^{N} \langle h_{i}^{n-s} \rangle$ can really be replaced with $N \langle h_{1}^{n-s} \rangle$ (where $h_1$ represents the first mound). Next, note that terms $\langle \sum h_{j_1}\text{...}h_{j_s} \rangle$ are really just the $s$-th moment of $h_\text{SD}$, $\langle \sum h_{j_1}\text{...}h_{j_s}\rangle = \langle h_\text{SD}^s\rangle$. This is true even though the indexes $ j_1\text{ ... }j_s$ are required to be different than $i$ since, if the domain size is infinite, the number of mounds, $N$, also tends to infinity, and any mean of the topography is the same whether we remove one mound or not. Making a substitution $j \equiv n-s$, we finally get the recurrence relation
\begin{linenomath*}
\begin{equation}\label{recur1}
\big\langle h_\text{SD}^n \big\rangle = N \sum_{j=1}^{n} {n-1 \choose n-j} \big\langle h_{1}^j \big\rangle \big\langle h_\text{SD}^{n-j} \big\rangle    \text{  . }
\end{equation}
\end{linenomath*}

To complete the proof, we only need to find the moments of individual mounds, $\langle h_{1}(\mathbf{x})^j \rangle$. For some quantity $g$ that depends on radius $r$ and position $\mathbf{x}_{0}$ the ensemble mean is given by
\begin{linenomath*}
\begin{equation}
\langle g \rangle = \int g(r,\mathbf{x}_{0}) f_r(r) f_{x_0}(\mathbf{x}_{0}) \text{d}r \text{d}^2\mathbf{x}_{0} \text{  , }
\end{equation}
\end{linenomath*}
where $f_r(r)$ is the exponential distribution of radii given by Eq. 2 of the main text and $f_{x_0}(\mathbf{x}_{0})$ is the uniform distribution of positions that is uniform over the domain, $f_{x_0}(\mathbf{x}_{0}) \equiv 1/L^2$, where $L$ is the domain size. We have that the $j$-th power of an individual mound is
\begin{linenomath*}
\begin{equation}
h_1(\mathbf{x})^j =  h_{m,0}^j \Big( \frac{r_1}{r_0} \Big)^j e^{-j (\mathbf{x}-\mathbf{x}_{0,1})^2 / 2 r_1^2} \text{  . }
\end{equation}
\end{linenomath*}
The $j$-th moment of an individual mound is then 
\begin{linenomath*}
\begin{equation}
\big\langle h_1^j \big\rangle = \int f_r(r_1) f_{x_0}(\mathbf{x}_{0,1}) h_{m,0}^j \Big( \frac{r_1}{r_0} \Big)^j e^{-j (\mathbf{x}-\mathbf{x}_{0,1})^2 / 2 r_1^2} \text{d}r_1 \text{d}^2\mathbf{x}_{0,1} \text{  . }
\end{equation}
\end{linenomath*}
Assuming that $L \rightarrow \infty$, we can first evaluate the integral over $\mathbf{x}_{0,1}$
\begin{linenomath*}
\begin{equation}
\int  f_{x_0}(\mathbf{x}_{0,1}) e^{-j (\mathbf{x}-\mathbf{x}_{0,1})^2 / 2 r_1^2} \text{d}^2\mathbf{x}_{0,1} = \frac{1}{L^2} \frac{2 \pi r_1^2}{j}  \text{  . }
\end{equation}
\end{linenomath*}
The term $ \frac{2 \pi r_1^2}{j}$ is simply the volume of a 2d Gaussian mound with a unit height and variance $r_1^2/j$. Next, we can evaluate the integral over $r_1$. We get
\begin{linenomath*}
\begin{equation}
\big\langle h_1^j \big\rangle = \frac{2 \pi h_{m,0}^j}{j} \frac{1}{L^2}  \int_0^\infty  \frac{r_1^{j+2}}{r_0^{j+1}} e^{-r_1/r_0} \text{d}r_1 = \frac{2 \pi h_{m,0}^j}{j} \frac{ r_0^2 }{L^2} \int_0^\infty  z^{2+j} e^{-z} \text{d}z = \frac{2 \pi h_{m,0}^j}{j} \frac{ r_0^2 }{L^2} \Gamma(3+j) \text{  , }
\end{equation}
\end{linenomath*}
where $\Gamma(n)$ is a gamma function. Using $\Gamma(n) = (n-1)!$ for integer $n$, we find
\begin{linenomath*}
\begin{equation}\label{individmoms}
\big\langle h_1^j \big\rangle = 2 \pi h_{m,0}^j \frac{ r_0^2 }{L^2} \frac{(2+j)!}{j} \text{  . }
\end{equation}
\end{linenomath*}
Moments of the individual mounds in Eq. \ref{recur1} are always multiplied by the number of mounds, $N$, and this product yields $N \langle h_1^j \rangle = 2 \pi \rho h_{m,0}^j \frac{(2+j)!}{j}$. Finally, returning to Eq. \ref{recur1}, we retrieve Eq. 5 of the main text
\begin{linenomath*}
\begin{equation}\label{momentsapp}
\big\langle h_\text{SD}^n \big\rangle = 2 \pi \rho \sum_{j=1}^{n} {n-1 \choose n-j}  \frac{(2+j)!}{j} h_{m,0}^j \big\langle h_\text{SD}^{n-j} \big\rangle    \text{  . }
\end{equation}
\end{linenomath*}
We note that high order moments of a numerical realization of our model on a domain of finite size do not conform to this prediction made under the assumption of infinite domain size. Namely, a finite-size domain necessarily has a finite maximum height, while the maximum height on an infinite domain would necessarily be infinite (see Fig. \ref{fig:Moment_prediction}). For this reason, when comparing high-order moments to real data, it is important to use the model of the same domain size and correlation length as the data. 

Equations 4 of the main text are simple consequences of the relations we derived here. In particular, the mean of the topography is 
\begin{linenomath*}
\begin{equation}
\big\langle h_\text{SD} \big\rangle = N \langle h_1 \rangle = 12 \pi \rho h_{m,0}   \text{  . }
\end{equation}
\end{linenomath*}
The variance is defined as $\sigma(h_\text{SD})^2 \equiv \langle h_\text{SD}^2 \rangle - \langle h_\text{SD} \rangle^2$. Using the recursion relation to find $\langle h_\text{SD}^2 \rangle$ and the relation for the mean we just derived, we get 
\begin{linenomath*}
\begin{equation}
\sigma(h_\text{SD}^2)^2  = 24 \pi \rho h_{m,0}^2   \text{  . }
\end{equation}
\end{linenomath*}

\subsubsection{Proof of Eq. 7 of the main text}

Here we derive Eq. 7 of the main text for the height correlation function of the synthetic ``snow dune'' topography. This proof follows very similar steps as the proof for the variance of the topography shown in the previous section. For this reason, we will only outline the proof and skip some of the details.

We start with the definition of the height correlation function given by Eq. 6 of the main text
\begin{linenomath*}
\begin{equation}\label{corr0}
 C_h(\mathbf{l}) \equiv  \frac{\langle h_\text{SD}(\mathbf{x})h_\text{SD}(\mathbf{x} + \mathbf{l}) \rangle - \langle h_\text{SD}\rangle^2}{ \sigma ^2(h_\text{SD}) } \text{  . } 
\end{equation}
\end{linenomath*}
Using the fact that $h_i$ and $h_j$ are independent for $i \neq j$, we can write the term $\langle h_\text{SD}(\mathbf{x})h_\text{SD}(\mathbf{x} + \mathbf{l}) \rangle$ as a sum
\begin{linenomath*}
\begin{equation}
\langle h_\text{SD}(\mathbf{x})h_\text{SD}(\mathbf{x} + \mathbf{l}) \rangle =  \sum_{i,j=1}^{N} \langle h_i(\mathbf{x}) h_j(\mathbf{x} + \mathbf{l}) \rangle = \sum_{i=1}^{N} \langle h_i(\mathbf{x})h_i(\mathbf{x} + \mathbf{l}) \rangle + \sum_{i\neq j}^N \langle h_i(\mathbf{x}) \rangle \langle h_j(\mathbf{x} + \mathbf{l})  \rangle  \text{  , } 
\end{equation}
\end{linenomath*}
Then, since the averages do not depend on which mound we are considering, and since $N \gg1$, we have
\begin{linenomath*}
\begin{equation} \label{corr1}
\langle h_\text{SD}(\mathbf{x})h_\text{SD}(\mathbf{x} + \mathbf{l}) \rangle  = N  \langle  h_1(\mathbf{x})h_1(\mathbf{x} + \mathbf{l})  \rangle + N(N-1) \langle h_1 \rangle^2 \approx N  \langle  h_1(\mathbf{x})h_1(\mathbf{x} + \mathbf{l})  \rangle +  \langle h_\text{SD} \rangle^2 \text{  , } 
\end{equation}
\end{linenomath*}
where we have used $N(N-1) \approx N^2$ and $N^2 \langle h_1 \rangle^2 =  \langle h_\text{SD} \rangle^2$ from the previous subsection. We, therefore, only need to find the term $\langle  h_1(\mathbf{x})h_1(\mathbf{x} + \mathbf{l})  \rangle$. Proceeding in the same way as in the previous subsection, we have 
\begin{linenomath*}
\begin{equation}
\langle  h_1(\mathbf{x})h_1(\mathbf{x} + \mathbf{l})  \rangle = \int f_r(r_1) f_{x_0}(\mathbf{x}_{0,1}) h_{m,0}^2 \Big( \frac{r_1}{r_0} \Big)^2 e^{- ( (\mathbf{x}-\mathbf{x}_{0,1})^2 + (\mathbf{x}  + \mathbf{l} -\mathbf{x}_{0,1})^2 ) / 2 r_1^2} \text{d}r_1 \text{d}^2\mathbf{x}_{0,1} \text{  . } 
\end{equation}
\end{linenomath*}
Reducing this integral into a manageable form involves similar steps as in the previous subsection - first perform an integration over $\mathbf{x}_{0,1} $ and then change from integration over $r_1$ to integration over $z = r_1/r_0$. We skip the detailed steps and simply show the result of these manipulations
\begin{linenomath*}
\begin{equation}
\langle  h_1(\mathbf{x})h_1(\mathbf{x} + \mathbf{l})  \rangle = \pi h_{m,0}^2 \frac{r_0^2}{L^2} \int_0^\infty z^4 e^{-z-l^2/(2r_0z)^2} \text{d} z \text{  . } 
\end{equation}
\end{linenomath*}
Returning this into Eq. \ref{corr1} and Eq. \ref{corr0}, and using the formula for the variance derived in the previous subsection, we retrieve Eq. 7 of the main text for the autocorrelation function. 

\subsection{Relationship between the ``snow dune'' topography and the gamma distribution}

\begin{figure}
\centering
\includegraphics[width=0.6\linewidth]{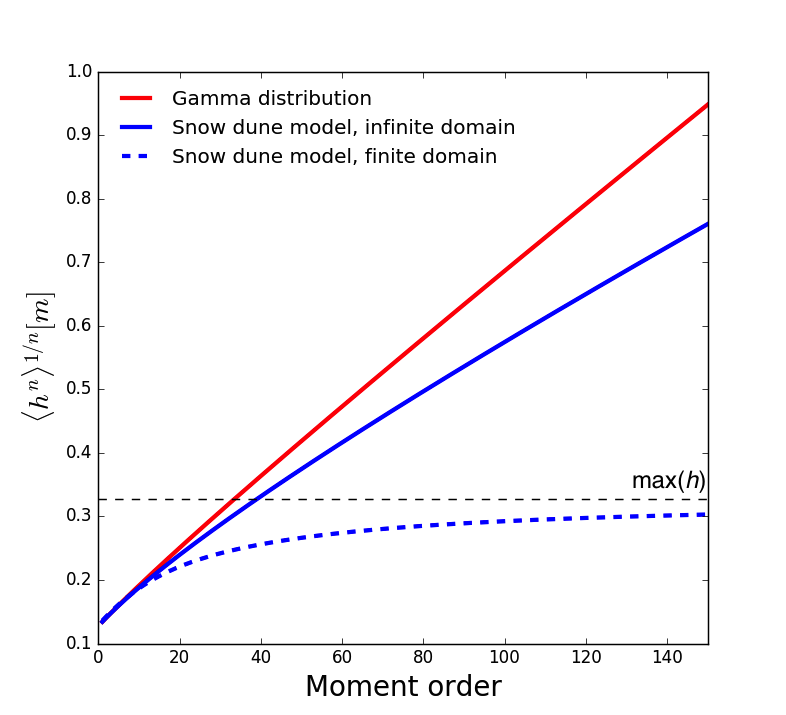}
\caption{First 150 root-moments for the gamma distribution (red line), theoretical predication for the ``snow dune'' model using Eq. \ref{momentsapp} that assumes an infinite domain size (solid blue line), and a numerical simulation of the ``snow dune'' model on a finite domain (dashed blue line). The maximum of the topography (horizontal black dashed line) is the asymptote for the root-moments of the finite-domain simulation.}
\label{fig:Moment_prediction}
\end{figure}

We will now use the results from the previous section to explore the relationship between the height distribution of the synthetic ``snow dune'' topography and the gamma distribution. We will show that the two distributions are not the same but are qualitatively similar. We compare the moments of these two distributions in Fig. \ref{fig:Moment_prediction}. 

The gamma distribution has moments
\begin{linenomath*}
\begin{equation}
\big\langle h^n \big\rangle_\Gamma = h_0^n \frac{\Gamma(m + k)}{\Gamma(k)} = h_0^n k (k+1)\text{...}(k+n-1)  \text{  , }
\end{equation}
\end{linenomath*}
where $\langle \text{...}\rangle_\Gamma$ is the mean with respect to the gamma distribution that has a scale parameter $h_0$ and a shape parameter $k$. We can rewrite this in a form in which we can directly compare the moments of the gamma distribution with the moments of the of the synthetic ``snow dune'' height distribution. In particular, we use the observation that $\langle h^n \rangle_\Gamma = h_0 (k+n-1)\langle h^{n-1} \rangle_\Gamma$ to write a recursion formula for the moments of the gamma distribution
\begin{linenomath*}
\begin{equation}
\big\langle h^n \big\rangle_\Gamma =  k \sum_{j=1}^{n} {n-1 \choose n-j} h_0^j \Gamma(j) \big\langle h^{n-j} \big\rangle_\Gamma    \text{  . }
\end{equation}
\end{linenomath*}
Comparing this equation with the recursion formula for the moments of the ``snow dune'' topography, Eq. \ref{recur1}, we can see that they have exactly the same form. In particular, in both cases $\langle h^n \rangle = \sum_{j} a_j {n-1 \choose n-j} \langle h^{n-j} \rangle$, for some coefficients $a_j$. The only difference is in the coefficients $a_j$ - for the ``snow dune'' topography, $a_j = N\langle h_{1}^j \rangle$, whereas for the gamma distribution $a_j = k h_0^j \Gamma(j)$. Expressing the moments of individual mounds, $\langle h_{1}^j \rangle$ using Eq. \ref{individmoms}, and identifying $\rho = k /(6\pi) $ and $h_{m,0} = h_0 / 2$ (Eqs. 11 of the main text), we find that for the ``snow dune'' topography $a_j = N\langle h_{1}^j \rangle = k h_{0}^j \frac{\Gamma(3+j)}{3 j 2^j}$. Therefore the coefficients $a_j$ in both distributions depend in the same way on parameters $k$ and $h_0$, and are only different in the numerical constants. Moreover, ``snow dune'' topography and gamma distribution have the same coefficients $a_1$ and $a_2$ by construction. 

The similarity between the two distributions is not unexpected - the gamma distribution is observed when the quantity of interest is a sum of $k$ exponentially distributed, independent random variables, $\sum_{i=1}^{k} h_i$. This means that the gamma distribution will necessarily have an expansion of the form Eq. \ref{recur1}, which was derived only under the assumption that the quantity we are interested in is a sum of independent random variables. In this recursion relation for the gamma distribution, the moments of individual terms, $\langle h_1^j \rangle$, are estimated with respect to the exponential distribution, $\langle h_1^j \rangle_\text{Exp} = h_0^j \Gamma(j)$. Therefore, the difference between the two distributions comes only from the fact that the height of individual mounds in the ``snow dune'' model is not exponentially distributed and therefore the moments $\langle h_1^j \rangle$ differ from those that enter the gamma distribution. The moments of individual mounds are, however, affected by arbitrary choices in our model such as the distribution of radii of individual mounds, and are therefore not robust.

\subsection{Sensitivity to model details}\label{sec:details_sensitivity}

\begin{figure}
\centering
\includegraphics[width=1.\linewidth]{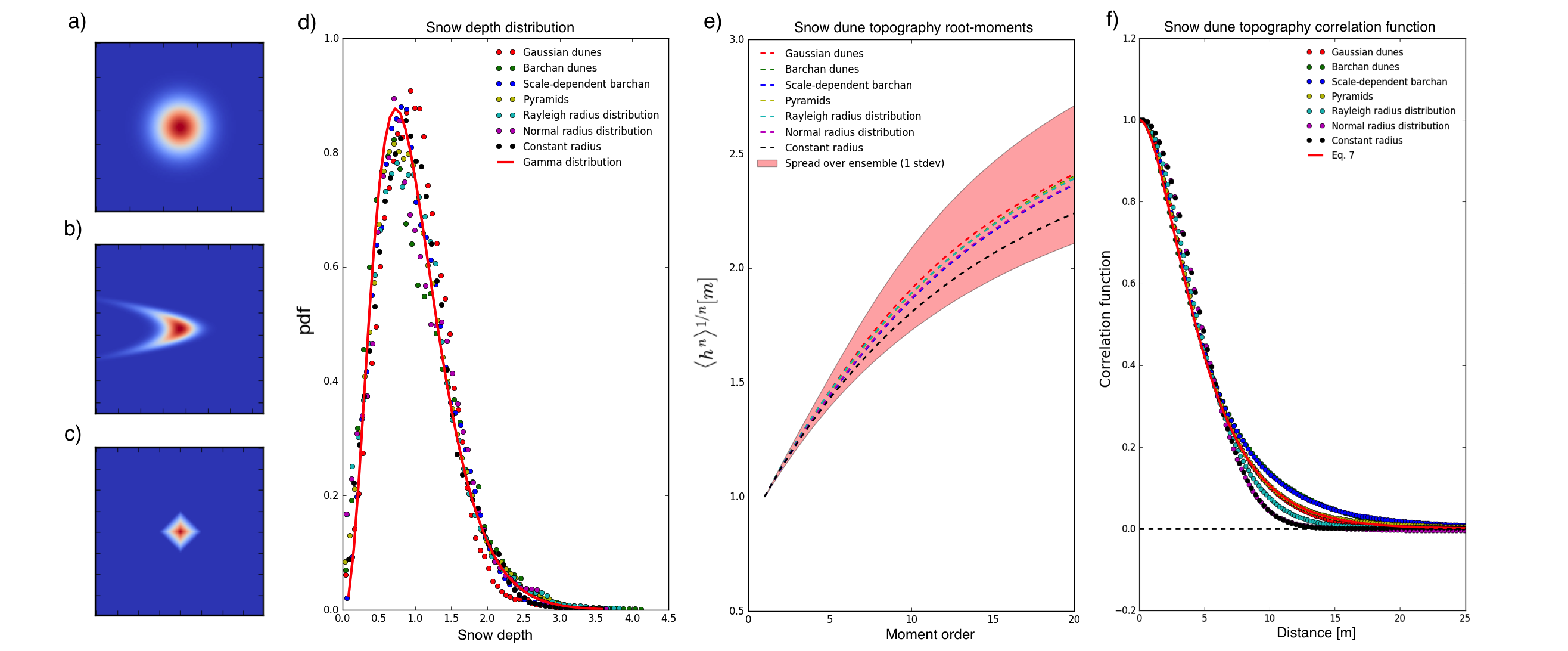}
\caption{a) A Gaussian mound. b) A barchan-like mound. c) A pyramid mound. d) Dots represent the snow depth distribution in the model assuming different mound shapes shown in panels a-c and different mound radii distributions. For cases when we did not use an exponential radius distribution, we assumed a Gaussian mound shape. For each topographic model, we chose $\rho$ and $h_{m,0}$ so that the mean snow depth is 1, the standard deviation is 0.5. The solid red line is the Gamma distribution with the same mean and standard deviation. e) The horizontal axis is the moment order, $n$, while the vertical axis is the $n$-th root of the $n$-th moment. The dashed lines represent the mean moments across an ensemble of 50 randomly-generated topographies with the same domain size and correlation length as the 2009N measurements. The colored shadings represent one standard deviation from the mean for the Gaussian dune case. f) Dots represent the radially averaged correlation function for varying model details. The solid red line is the analytical prediction for Gaussian dunes using Eq. 7 of the main text. For each topographic model, we chose $r_0$ so that the correlation length is the same as for the 2009N measurements.}
\label{fig:dune_shapes}
\end{figure}

We saw in the previous sections that simply due to the fact that our topography is a sum of many uncorrelated dunes, the moments of the height distribution follow the universal recursion relation, Eq. \ref{recur1}. Moments of individual mounds, $\langle h_1^n \rangle$, enter this recursion relation, so changing the model details such as the shapes of the mounds or their distribution of radii in principle change the height distribution of the topography. The same is true for the correlation function. However, it is unclear whether these effects are important and detectible on the scales for which we have measurements. Here, we test the sensitivity of our model to such details.  

Besides the standard Gaussian mounds, we also considered models using mounds of the following form
\begin{linenomath*}
\begin{gather}
h(\mathbf{x}) = h_{m} e^{- \frac{1}{2} \left( \frac{x-x_0}{r} \right)^2 - \frac{1}{2} \left( \frac{y-y_0}{r} + \left( \frac{x-x_0}{r} \right)^2 \right)^2 } \quad \text{ , } \quad \text{barchan-like} \label{barchan}\\
h(\mathbf{x}) = h_{m} \text{max}\left( 0, 1- \left(\left|\frac{x-x_0}{r}\right| + \left|\frac{y-y_0}{r}\right|\right)  \right) \quad \text{ , } \quad \text{pyramid} \label{pyramid}
 \text{  . }
\end{gather}
\end{linenomath*}
These three different mound shapes are shown in Fig. \ref{fig:dune_shapes}a-c. We note that the anisotropic barchan-like mounds, described by Eq. \ref{barchan} and shown in Fig. \ref{fig:dune_shapes}b, are loosely motivated by real barchan dunes \citep{filhol2015snow} but we did not compare Eq. \ref{barchan} to real barchans. In addition to this, we also considered ``scale-dependent barchans'' that have a Gaussian shape if $r \ll r_0$, and a barchan-like shape if $r \gg r_0$, with a smooth transition between the two. As in the original model, we assumed the the mound height is proportional to its horizontal scale, $h_m = h_{m,0} r/r_0$, for all models. Besides the exponential mound radius distribution we considered in the main text, in the case of Gaussian mounds, we also considered a normal distribution with standard deviation equal to 0.5 of the mean, a Rayleigh distribution, and a case where all mounds have the same radius, $r \equiv r_0$. For non-Gaussian mounds, we assumed an exponential mound radius distribution in all cases.

In all cases, the relationships between model parameters $h_{m,0}$, $\rho$, and $r_0$ and measurable statistics of the topography $\langle h \rangle$, $\sigma(h)$, and $l_0$ have the same form as for the original model (Eqs. 4 and 8 of the main text), but the constants of proportionality are different. Thus, to compare the models, in each case we chose $h_{m,0}$, $\rho$, and $r_0$, so that all models have the same measurable statistics $\langle h \rangle$, $\sigma(h)$, and $l_0$. Moreover, we chose the ratio of domain size and correlation length to be the same as for 2009N measurements in order to ascertain whether the differences between the models are detectible on the scale of these measurements. 

In Fig. \ref{fig:dune_shapes}d, we show the snow depth distribution for these models and compare it to a gamma distribution with parameters calculated using  $\langle h \rangle$ and $\sigma(h)$ used in the models. We can see that the all of these distributions agree well with each other to within a spread induced by a finite domain size. Next, in Fig. \ref{fig:dune_shapes}e, we look at the root-moments of the distributions for these models. We can see that all of them fall well within the 1-$\sigma$ uncertainty range of each other, which means that they are most likely indistinguishable on the scale of the measurements. Finally, in Fig. \ref{fig:dune_shapes}f, we compare the radially averaged height correlation function, $C_h(l)$, between these models. Here, some deviation from Eq. 7 can be observed for different models at $l$ larger than $l_0$ and smaller than several times $l_0$, but this deviation is modest (largest deviation of about 0.1 for constant radius mounds of about). 

We conclude that the snow-depth distribution is insensitive to mound shape and radius distribution, at least on the scale at which the measurements were taken. We expect that this insensitivity holds on much larger scales, and even in the limit of infinite domain size, model details would only become significant for high-order moments. Correlation function is observably different, but these differences are small. We have not explicitly tested the model sensitivity to the linear relationship between mound height and radius, it is not yet clear how important this model assumption is. Nevertheless, we can conclude that our model is insensitive to most model details which explains why we find such a good agreement with the LiDAR measurements.

\section{Ice heat conduction}\label{sec:thick_evol}

In this section, we will derive Eqs. 12 to 16 for the heat flux conducted through the ice under snow cover that is well-described with our ``snow dune'' model. We will also consider heat conduction for arbitrary parameters $\eta$, $\Sigma$, and $\Lambda$

\subsection{Non-dimensional heat equation}

First, we show that the conductive heat flux must be of the form stated in Eq. 12 of the main text, $F_c = F_0 \Phi(\eta,\Sigma,\Lambda)$.

As we stated in the main text, we are assuming that ice is a block of uniform thickness, $H$, that ice and snow have fixed conductivities, $k_i$ and $k_s$, that the snow cover is well-described with our ``snow dune'' model, that the temperature at the ice-ocean interface is fixed at the freezing point of salt water, $T_f$, that the temperature of the snow surface is fixed at the temperature of the atmosphere, $T_a$, and that the temperature field, $T$, within the ice and snow is in a steady state. Finally, we assume periodic boundary conditions in the horizontal. We define the ice-ocean interface to be the plane $z = -H$, and the snow-ice interface to be the plane $z=0$.  These assumptions are described by the following set of equations
\begin{linenomath*}
\begin{gather}\label{heat_equation}
\nabla^2 T = 0 \quad \text{ , } \quad T\big(z = -H\big) = T_f  \quad \text{ , } \\
 T\big(z = h_\text{SD}(x,y)\big) = T_a \quad \text{ , } \quad k_i\frac{\partial T}{\partial z}\big(z = 0^-\big) = k_s\frac{\partial T}{\partial z}\big(z = 0^+\big) \text{  , }
\end{gather}
\end{linenomath*}
where $\nabla^2\equiv \frac{\partial^2 }{\partial x^2} +  \frac{\partial^2 }{\partial y^2} +  \frac{\partial^2 }{\partial z^2}$ is the Laplacian, $x$ and $y$ are the horizontal coordinates, $z$ is the vertical coordinate, and $h_\text{SD}(x,y)$ is the snow depth. The last equation represents the condition that the heat flux is continuous across the snow-ice interface. 

At this point we make another simplifying assumption that we stated in the main text - we assume that, within snow, heat is transported purely vertically so that, at steady state and assuming a constant snow conductivity, the temperature profile is linear, $T(z>0) = T_a + \big(\frac{z}{h_\text{SD}(x,y)} - 1\big)\big( T_a - T(z=0)\big)$. With this simplification, we only need to solve the heat equation within the rectangular ice domain. In this case, the upper-surface boundary condition becomes $k_i\frac{\partial T}{\partial z}(z = 0) = k_s\frac{T_a - T(z=0) }{h_\text{SD}(x,y)}$. Next, we non-dimensionalize the variables in the above equations as 
\begin{linenomath*}
\begin{equation}
\theta \equiv \frac{T-T_f}{T_a-T_f} \quad \text{ , } \quad \tilde{x} \equiv \frac{x}{l_0} \quad \text{ , } \quad \tilde{y} \equiv \frac{y}{l_0} \quad \text{ , } \quad \tilde{z} \equiv \frac{z}{H} \quad \text{ , } \quad \tilde{h} \equiv \frac{h_\text{SD}}{\langle h_\text{SD} \rangle} \text{  . }
\end{equation}
\end{linenomath*}
In terms of these non-dimensional variables and non-dimensional parameters, $\eta = \frac{k_i}{k_s}\frac{\langle h \rangle}{H}$, $\Sigma = \frac{\sigma(h)}{\langle h \rangle}$, and, $\Lambda = \frac{l_0}{H}$, introduced in Eq. 13 of the main text, we can rewrite the heat equation as 
\begin{linenomath*}
\begin{equation}
\frac{1}{\Lambda^2} \Big( \frac{\partial^2 \theta}{\partial \tilde{x}^2} +\frac{\partial^2 \theta}{\partial \tilde{y}^2} \Big) +  \frac{\partial^2 \theta}{\partial \tilde{z}^2}  = 0 \quad \text{ , } \quad \theta \big(\tilde{z} = -1\big) = 0  \quad \text{ , } \quad \frac{\partial \theta}{\partial \tilde{z} }\big(\tilde{z} = 0\big) = \frac{1-\theta(\tilde{z} = 0)}{\eta \tilde{h} } \text{  . }  \label{nondim_heat}
\end{equation}
\end{linenomath*}
Therefore, the non-dimensional temperature, $\theta$, can only depend on $\eta$, $\Lambda$, and the parameters that determine the non-dimensional snow surface, $\tilde{h}$. If the snow surface is well-described by our ``snow dune'' model, the statistics of the non-dimensional snow topography, $\tilde{h}$, can only depend on $\Sigma$, since the mean is fixed to 1 and the typical dune radius, $r_0$, is fixed by rescaling the horizontal coordinates, $x$ and $y$, by $l_0$. So, if our ``snow dune'' model applies, the spatially averaged statistics of $\theta$ can only depend on $\eta$, $\Lambda$, and $\Sigma$.

Heat flux through any surface characterized by a normal vector, $\mathbf{n}$, within the ice can be calculated as $F_{\mathbf{n}} = k_i\nabla T \cdot \mathbf{n}$. In particular, flux through the snow-ice interface is $k_i \partial T/\partial z (z=0)$. In terms of non-dimensional temperature, the flux is then $F_0 \partial \theta/\partial \tilde{z}(\tilde{z} = 0)$, where $F_0 \equiv k_i \frac{T_f-T_a}{H}$, as in Eq. 13. The mean flux through the snow-ice interface is then 
\begin{linenomath*}
\begin{equation}
F_c = F_0 \Big\langle \frac{\partial \theta}{\partial \tilde{z}} (\tilde{z} = 0) \Big\rangle  \text{  , }
\end{equation}
\end{linenomath*}
where $\langle \text{...}\rangle$ stands for averaging along the horizontal. Since $\langle \partial \theta/\partial \tilde{z} (\tilde{z} = 0) \rangle$ can only depend on $\eta$, $\Lambda$, and $\Sigma$, we have
\begin{linenomath*}
\begin{equation}
F_c = F_0 \Phi\big(\eta,\Sigma,\Lambda\big)  \text{  , }
\end{equation}
\end{linenomath*}
and we recover Eq. 12.

\subsection{Limiting behavior}

We now solve the non-dimensional heat equation for the limits $\Sigma \rightarrow 0$, $\Lambda \rightarrow \infty$, and $\Lambda \rightarrow 0$. These limits yield non-dimensional heat flux assuming a uniform snow cover, $\Phi_u$, a purely vertical heat transport, $\Phi_v$, and a dominantly horizontal heat transport, $\Phi_h$. 
\begin{enumerate}
\item In the limit of $\Sigma \rightarrow 0$, the snow cover is uniform and $\tilde{h} = 1$ everywhere. In this case, heat conduction is purely vertical and the temperature profile is linear, $\theta = \theta_0(\tilde{z} + 1)$, where $\theta_0$ is the non-dimensional temperature at the snow-ice interface and we used the bottom boundary condition, $\theta(\tilde{z} = -1)=0$. We obtain $\theta_0$ from the top boundary condition, $\theta_0 = \frac{1-\theta_0}{\eta}$, and thus find $\theta_0 =  \frac{1}{1+\eta}$. With that, the non-dimensional flux through the ice under a uniform snow cover is $\Phi_u = \frac{\text{d} \theta}{\text{d} \tilde{z}} = \theta_0$. We thus have
\begin{linenomath*}
\begin{equation}\label{uniform}
\Phi_u = \frac{1}{1+\eta}  \text{  . }
\end{equation}
\end{linenomath*}

\item In the limit $\Lambda \rightarrow \infty$, the horizontal extent of the snow features is much larger than the thickness of the ice. So, heat will again be transported vertically, but the temperature profile within the ice will be set by the local snow height. In other words, the temperature profile will be $\theta = \theta_0\big( \tilde{x}, \tilde{y} \big) (\tilde{z} + 1)$, where $\theta_0\big( \tilde{x}, \tilde{y} \big)$ is the non-dimensional temperature at the snow-ice interface at the point $\tilde{x}, \tilde{y}$. As in the previous case, we find $\theta_0\big( \tilde{x}, \tilde{y} \big)$ from the top boundary condition at $\tilde{x}, \tilde{y}$, $\theta_0\big( \tilde{x}, \tilde{y} \big) =  \frac{1}{1+\eta\tilde{h}(\tilde{x}, \tilde{y})}$. Since the local flux is $ \frac{\text{d} \theta}{\text{d} \tilde{z}} = \theta_0\big( \tilde{x}, \tilde{y} \big)$, we find 
\begin{linenomath*}
\begin{equation}\label{Fv}
\Phi_v = \Big\langle \frac{\text{d} \theta}{\text{d} \tilde{z}}(\tilde{z} = 0) \Big\rangle = \Big\langle \frac{1}{1+\eta\tilde{h}} \Big\rangle = \int_0^\infty \frac{f_\Gamma(\tilde{h})}{1+\eta\tilde{h}} \text{d}\tilde{h} \text{  , }
\end{equation}
\end{linenomath*}
where we used the fact that the height distribution of our ``snow dune'' topography is well-described with a gamma distribution. Since the non-dimensional snow topography, $\tilde{h}$, has a mean equal to 1 and a standard deviation equal to $\Sigma$, the parameters of this gamma distribution are, according to Eq. 10 of the main text, $k = \Sigma^{-2}$ and $h_0 = \Sigma^2$. Thus, we recover Eq. 15 of the main text for the purely vertical heat transport.

\item In the limit $\Lambda \rightarrow 0$, the horizontal extent of the snow features is much smaller than the thickness of the ice. In this case, temperature must be uniform in the horizontal, i.e. $\theta = \theta(\tilde{z})$. In the vertical, the temperature profile must again be linear, since the mean heat flux across any horizontal plane within the ice must be constant if the temperature is in a steady state. Thus, again, the temperature must be of the form  $\theta = \theta_0(\tilde{z} + 1)$. However, in this case, the snow depth is variable, and, if $\theta_0$ is constant in the horizontal, the top boundary condition cannot be met at every point along the snow-ice interface. However, we can still determine $\theta_0$ by requiring that the total flux conducted across the snow-ice interface remains constant. At each point along the snow-ice interface, the local non-dimensional flux through the snow is $\frac{1-\theta_0}{\eta \tilde{h}(\tilde{x},\tilde{y})}$. So, the mean non-dimensional flux through the upper boundary is $\frac{1-\theta_0}{\eta} \big\langle \frac{1}{\tilde{h}}\big\rangle$. On the other hand, using $\theta = \theta_0(\tilde{z} + 1)$, this flux must be equal to $ \frac{\text{d} \theta}{\text{d} \tilde{z}} = \theta_0$. Thus, equating the two fluxes, $\theta_0 = \frac{1-\theta_0}{\eta} \big\langle \frac{1}{\tilde{h}}\big\rangle$, we find $\theta_0 = \frac{\langle \tilde{h}^{-1}\rangle}{\langle \tilde{h}^{-1}\rangle + 1}$. Thus, the non-dimensional flux in the limit $\Lambda \rightarrow 0$ is 
\begin{linenomath*}
\begin{equation} \label{Fh1}
\Phi_h = \frac{\big\langle \tilde{h}^{-1}\big\rangle}{\big\langle \tilde{h}^{-1}\big\rangle + 1} \text{  . }
\end{equation}
\end{linenomath*}
If the snow-depth distribution is well-described with a gamma distribution with a shape parameter $k$, the average $\langle \tilde{h}^{-1}\rangle$ is equal to $\frac{k-1}{k}$ for $k \geq 1$. Since, in our case, $k = \Sigma^{-2}$, we have 
\begin{linenomath*}
\begin{equation}\label{horizontal_supp}
\Phi_h = \frac{1}{1 + \eta(1-\Sigma^2)} \text{  , }
\end{equation}
\end{linenomath*}
and we recover Eq. 16 of the main text. For $k \leq 1$, $\langle \tilde{h}^{-1}\rangle$ is infinite, and Eq. \ref{horizontal_supp} is no longer valid. Nevertheless, Eq. \ref{Fh1} is still valid and implies that $\Phi_h = 1$ for $k > 1$, meaning that the heat flux is the same as if there were no snow on top of the ice. Clearly, for $\Lambda = 0$, we could not satisfy the upper boundary condition at every point along the interface. Nevertheless, for any small but non-zero $\Lambda$, a boundary layer of thickness on the order $\Lambda$ develops within the ice and near the snow-ice interface. Within this boundary layer, vertical heat transport dominates, and the temperature varies along the horizontal to match the upper boundary condition at every point.
\end{enumerate}

\subsection{Finite $\Lambda$}

Finally, we derive Eq. 14 of the main text for the non-dimensional flux $\Phi$ assuming an arbitrary $\Lambda$. To understand how the heat transport changes as we increase $\Lambda$ from 0 to $\infty$, we numerically solved the non-dimensional heat conduction problem, Eqs. \ref{nondim_heat}, on a 50x50x20 grid using a finite volume method. We set the typical mound radius, $r_0$, to be 1 grid point so that the total domain is roughly 5 correlation lengths wide. For each set of parameters $\eta$ and $\Sigma$, we sweeped the parameter $\Lambda$ from $\Lambda = 0.01$ to $\Lambda = 5$. Moreover, for each combination of $\eta$, $\Sigma$, and $\Lambda$ we solved the heat conduction problem for 100 random realizations of the ``snow dune'' topography. For each of these realization, we calculated the mean heat flux across the domain and then found the mean flux across these 100 realizations. We chose to do this rather than to increase the grid size since our numerical scheme becomes increasingly inefficient for even modest increases in grid size. 

The numerical scheme is especially inefficient at small $\Lambda$. This is a problem, since, for small $\Lambda$, high vertical resolution is required to resolve the boundary layer near the snow-ice interface we discussed in the previous subsection. We did not attempt to resolve this issue in a principled way, and simply carried out our low-resolution estimates. However, to ensure discrepancies at small $\Lambda$ between the numerical results and the theoretical prediction given by Eq. \ref{horizontal_supp} are due to low resolution, we carried out one trial run at $\Lambda =0.01$ (with $\eta = 1$ and $\Sigma = 0.75$) where we kept the snow topography fixed with a horizontal resolution of 25x25 and progressively increased the vertical resolution from 25 to 800 grid points. We found that with the vertical resolution of 25 grid points, the flux from numerical calculations was around $0.94 \Phi_h$, and became closer and closer to $\Phi_h$ with increasing resolution. At a resolution of 800 vertical grid points the numerical flux was around $0.993 \Phi_h$ found using Eq. \ref{Fh1}. 

\begin{figure}
\centering
\includegraphics[width=1.\linewidth]{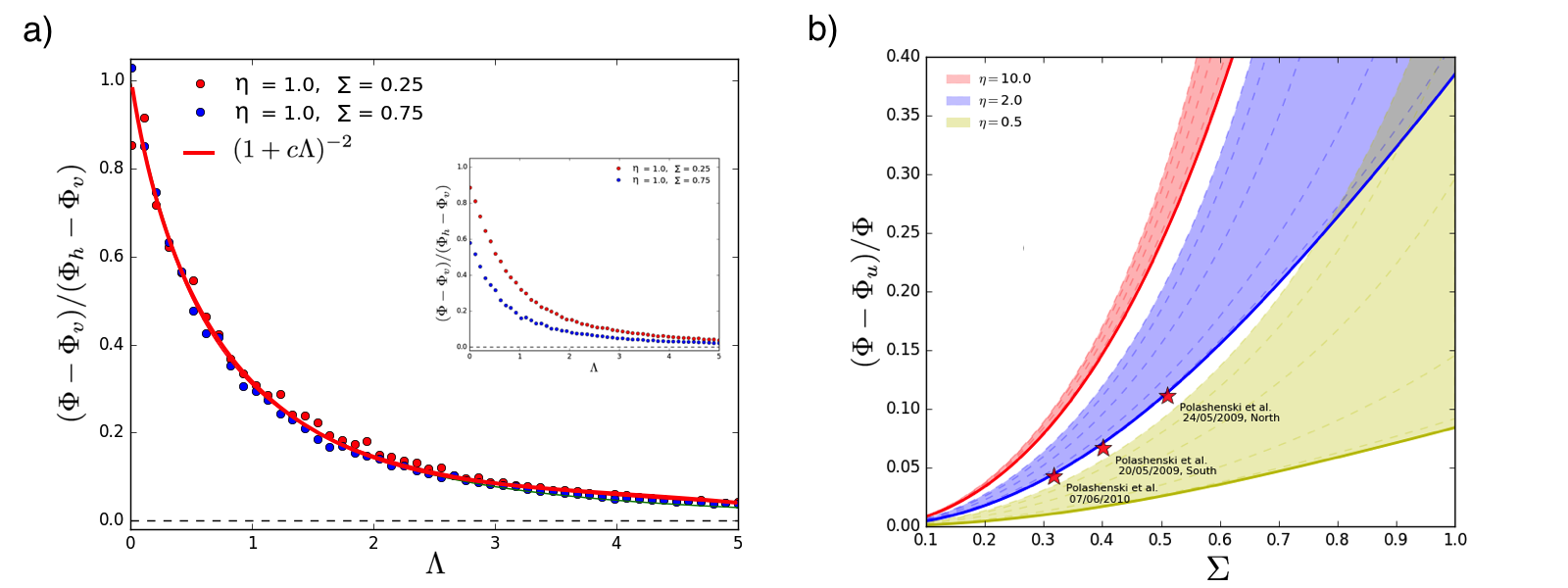}
\caption{a) The ratio $\varphi = \frac{\Phi-\Phi_v}{\Phi_h - \Phi_v}$ as a function of the non-dimensional parameter, $\Lambda$. Each dot represents the average $\varphi$ across an ensemble of 100 runs with different random realizations of the ``snow dune'' topography and the same values of non-dimensional parameters. Since the simulated ratio $\varphi$ does not tend to 1 (see the inset) as $\Lambda$ tends to 0 due to low resolution of our simulations, the simulations are normalized so that $\varphi(\Lambda = 0) = 1$. The solid red curve represents the function $(1+c\Lambda)^{-2}$ with $c \approx 0.83$. Inset shows the simulation results before normalizing to $\varphi(\Lambda = 0) = 1$. b) The fraction of the flux due to snow-depth variability, $\frac{\Phi - \Phi_u}{\Phi}$ as a function of non-dimensional snow roughness $\Sigma$. The flux $\Phi$ is calculated using Eq. \ref{total_supp} and $\Phi_u$ is calculated using Eq. \ref{uniform}. Different colors stand for different non-dimensional depth, $\eta$. Thick colored lines represent the heat flux with purely vertical heat transport ($\Lambda \rightarrow \infty$), while the colored shadings represent the flux attainable when the horizontal heat diffusion is included ($\Lambda < \infty$). Thin dashed lines represent the heat flux for $\Lambda$ equal to 8, 2, 0.5, 0.125, and 0. The stars represent the fraction of the flux due to snow variability under the measured LiDAR topographies of \citet{polashenski2012mechanisms} with parameters given in Table II of the main text.}
\label{fig:finite_lambda}
\end{figure}

Since we know that in the limit $\Lambda \rightarrow 0$ the non-dimensional flux must approach $\Phi_h$ and that in the limit $\Lambda \rightarrow \infty$, it must approach $\Phi_v$, we consider the ratio $\varphi \equiv \frac{\Phi-\Phi_v}{\Phi_h - \Phi_v}$, where $\Phi$ is the numerical flux, calculated at finite $\Lambda$. This ratio must decrease from 1 at $\Lambda = 0$ to 0 for large $\Lambda$. We plot this ratio in Fig. \ref{fig:finite_lambda}a. The simulated $\varphi$, however, does not reach 1 at small $\Lambda$ (see inset of Fig. \ref{fig:finite_lambda}a). In our single trial where we progressively increased the vertical resolution, this ratio was roughly 0.73 at a resolution of 25 vertical grid points, consistent with $\varphi$ in the inset of Fig. \ref{fig:finite_lambda} we obtained in our low resolution runs. The ratio increased with increasing resolution, reaching roughly 0.97 at a resolution of 800 vertical grid points, thus consistent with Eqs. \ref{Fv} and \ref{horizontal_supp} being valid. For this reason, we believe that our ensemble runs are at odds with predictions mainly due to low resolution of our simulations. Thus, we normalized the simulation results so that $\varphi(\Lambda = 0) = 1$. This is shown in Fig. \ref{fig:finite_lambda}a. We can see that in this case, the simulation results for both values of the parameters $\eta$ and $\Sigma$ roughly fall on the same curve that is well-fit with a function $(1+c\Lambda)^{-2}$ where $c \approx 0.83$. With this, we get
\begin{linenomath*}
\begin{equation}\label{total_supp}
\Phi \big( \eta, \Sigma, \Lambda \big) = \Phi_v + \frac{\Phi_h - \Phi_v}{(1+c\Lambda)^{2}}\text{  , }
\end{equation}
\end{linenomath*}
and we recover Eq. 14 of the main text.



\subsection{Exploring the parameter space}

Even though the horizontal heat transport does not contribute significantly to heat conduction on flat ice, it may still be of interest to understand how it might contribute in principle. Here, we use Eq. \ref{total_supp} to briefly explore how the heat flux behaves when changing parameters $\eta$, $\Sigma$, and $\Lambda$. We will focus our discussion on the fraction of the heat flux that is due to snow-depth variability, $\frac{\Phi - \Phi_u}{\Phi}$, where we find $\Phi$ using Eq. \ref{total_supp} and we find $\Phi_u$ using Eq. \ref{uniform}.

Our main results here are summarized in Fig. \ref{fig:finite_lambda}b. We find that the amount of flux that can be attributed to snow-depth variability strongly increases with the snow roughness, $\Sigma$, and also increases with the non-dimensional snow depth, $\eta$. We find that it is roughly proportional to $\Sigma^2$, although this dependence is highly approximate. The sensitivity to other parameters also increases with $\Sigma$. The effect of horizontal heat transport may be be large in principle: e.g. decreasing $\Lambda$ from $\sim 10$ to $0.1$ would roughly double the effect of snow variability on ice with the same $\eta$ and $\Sigma$ as 2009N LiDAR measurements. This is even more pronounced at larger $\Sigma$ and smaller $\eta$. As we have discussed in the previous subsections and the main text, for $\Sigma \geq 1$, there exists a possibility that snow is not registered at all if $\Lambda$ is small enough. This occurs because, if $\Sigma \geq 1$ a certain fraction of the ice is snow-free, so that if horizontal heat transport is efficient enough, all of the heat is exported through these regions. 

\section{Deriving stage I pond evolution} \label{sec:substageI}

In this section, we derive Eqs. 17 to 19 of the main text for pond evolution during stage I. We also develop a higher order approximation for pond coverage evolution and we test these approximations against a 2d model of pond evolution. We use the notation $\dot{x} \equiv \frac{\text{d} x}{\text{d} t}$ for the rate of change of a quantity $x$, as in section VI of the main text. 

\subsection{Proof of Eqs. 17 and 18 of the main text}

As we stated in the main text, we are assuming that snow is fully permeable, that there exists a common water table below which the snow is completely saturated with water, that underlying ice is impermeable and initially flat, that the initial snow topography is well-described by our ``snow dune'' topography, that no meltwater is lost from the domain, that snow, bare ice, ponded ice, and ponded snow (water-covered snow) melt at different rates that are constant in space and time, and that both snow and ice melt only at the surface.

We denote the ice and snow topography as $h(\mathbf{x})$, which represents the vertical coordinate of snow, or ice if there is no snow on top. Because we have assumed that snow is permeable, there exists a common water level, $w$, such that all points with $h(\mathbf{x}) < w$ are ponded. Since we have assumed that the underlying ice is initially flat, there also exists an initial common ice level, which we can define as $h = 0$, such that all points with $h(\mathbf{x}) < 0$ have no snow on top of ice. Because of this assumption, the depth of snow in snow-covered regions, $h(\mathbf{x}) > 0$, is given by $h(\mathbf{x})$. Since we have assumed that meltwater cannot be lost during stage I, pond coverage can only grow during this stage. We can see from the assumptions above that there can be no regions of bare ice (ice that is snow-free and pond-free). Therefore, it is enough to consider the melt of bare snow, ponded snow, and ponded ice. 

Since we have defined snow-free ice regions to be $h(\mathbf{x}) < 0$ and pond regions to be $h(\mathbf{x}) < w$, where $w$ is the height of the water table, we have
\begin{linenomath*}
\begin{gather} 
p_i = \int_{-\infty}^0 f(h,t) \text{d}h \text{  , } \label{pandpi} \\
p = \int_{-\infty}^w f(h,t) \text{d}h  \label{pandpi_p} \text{  , }
\end{gather} 
\end{linenomath*}
where $p_i$ is the ice fraction, $p$ is the pond fraction, and $f(h,t)$ is the surface height distribution that, due to different regions melting at different rates, depends on time, $t$. Since, under our assumptions, there are no regions of bare ice, bare snow fraction can be found as $1-p$ and ponded snow fraction as $p-p_i$. 

By knowing the melt rates of different regions of the ice, we can find the rate of change of the distribution $f$. Let us, for the moment, assume that the ice melt rate $\dot{h}$ is an arbitrary function of height, $\dot{h} = \dot{h}(h)$. We are interested in finding the the quantity $f(h,t + \text{d}t)\text{d}h$, i.e., the fraction of the ice surface with a height between $h$ and $h+\text{d}h$ at a time $t + \text{d}t$. To first order in $\text{d}t$, the surface with a height $h$ at time $t + \text{d}t$ used to be at a height $h - \dot{h}(h) \text{d}t$, while the surface at with a height $h + \text{d}h$ at time $t + \text{d}t$ used to be at a height $h + \text{d}h - \dot{h}(h + \text{d}h) \text{d}t$, where we are assuming a sign convention $\dot{h} < 0$. So, we have 
\begin{linenomath*}
\begin{equation}\label{fdt1}
f(h,t+\text{d}t)\text{d}h = f(h - \dot{h}\text{d}t, t)\text{d}h_2 \text{  , }
\end{equation}
\end{linenomath*}
for some height range $\text{d} h_2$ that may be different than the range $\text{d}h$ because the melt rate $\dot{h}(h)$ may be different than the melt rate $\dot{h}(h + \text{d}h)$. In particular, $\text{d}h_2$ is the difference between the heights $h + \text{d}h - \dot{h}(h + \text{d}h) \text{d}t$ and $h - \dot{h}(h) \text{d}t$, so to first order in $\text{d}t$, we have 
\begin{linenomath*}
\begin{equation} \label{dh2}
\text{d}h_2 = \text{d} h (1 - \text{d}t \frac{\text{d}\dot{h}}{\text{d}h}) \text{  . }
\end{equation}
\end{linenomath*}
Combining Eqs. \ref{fdt1} and \ref{dh2}, we find 
\begin{linenomath*}
\begin{equation}\label{fdt2}
f(h,t+\text{d}t) = f(h - \dot{h}\text{d}t, t)(1 - \text{d}t \frac{\text{d}\dot{h}}{\text{d}h}) \text{  . }
\end{equation}
\end{linenomath*}
Expanding this equation to first order in $\text{d} t$, we find the conservation law for the height probability distribution
\begin{linenomath*}
\begin{equation} \label{dfdt}
\frac{\partial f}{\partial t} = -  \frac{\partial (f \dot{h})}{\partial h}\text{  . }
\end{equation}
\end{linenomath*}
Note that, because ice is melting ($\dot{h} < 0$), it is the melt rate slightly above $h$ that enters the conservation law, Eq. \ref{dfdt}. This distinction will be important when considering the rates of change of coverage fractions due to the discontinuous change in melt rates when the surface type changes.

We can now express the rate of change of the pond coverage fraction as 
\begin{linenomath*}
\begin{equation} \label{pdot1}
\dot{p} = \frac{\text{d} }{\text{d} t} \int_{-\infty}^w f(h,t) \text{d}h = f(w,t) \Big(\dot{w} - \dot{h}_s\Big) \text{  , }
\end{equation}
\end{linenomath*}
where we have used Eqs. \ref{pandpi_p} and \ref{dfdt} and the fact that, since $w > 0$, the melt rate slightly above the water level, $w$, is the snow melt rate, $\dot{h}(w + \text{d} h) =  \dot{h}_s$. Since, under our assumptions, all non-ponded regions melt at the same constant rate, the height distribution for $h > w$ does not change shape and is simply shifted downwards as the time progresses, and we can express $f(w,t) = f(w - \dot{h}_s t)$. At time $t=0$, this distribution is well-described by the gamma distribution, so we find
\begin{linenomath*}
\begin{equation} \label{pdot2}
\dot{p} = f_\Gamma(w - \dot{h}_st) \Big(\dot{w} - \dot{h}_s\Big) \text{ . }
\end{equation}
\end{linenomath*}
Therefore, to close this equation, we only need to express $\dot{w}$.


The water level can be expressed in terms of the water volume, $V_\text{w}$. Because snow is porous, some fraction of the water is stored in the saturated snow. We consider all snow below $w$ to be saturated and all of the pores to be filled with water. For this reason, $1\text{ m}^3$ of saturated snow will contain $1-r_s$ meters cubed of water, where $r_s = \frac{\rho_s}{\rho_i}$. The volume of water can then be expressed as 
\begin{linenomath*}
\begin{equation} \label{Vw1}
V_\text{w} = V_\text{pi} + V_\text{ps} + V_\text{bs}\text{ , }
\end{equation}
\end{linenomath*}
where $V_\text{pi}$, $V_\text{ps}$, and $V_\text{bs}$ are the volumes of water contained in regions of ponded ice, ponded snow, and bare snow. A small element of area, $\Delta A$, of each of these categories then contains the following amounts of water
\begin{linenomath*}
\begin{gather} \label{dVs}
\Delta V_\text{pi} = (w-h(\mathbf{x}) ) \Delta A \text{  , } \\
\Delta V_\text{ps} = \Big( (w-h(\mathbf{x}) ) + h(\mathbf{x})(1-r_s) \Big) \Delta A = (w-r_s h(\mathbf{x}))  \Delta A  \text{  , } \\
\Delta V_\text{bs} = w(1-r_s) \Delta A \text{  . } 
\end{gather}
\end{linenomath*}
We can then find the total volumes of water in these regions as
\begin{linenomath*}
\begin{gather} \label{Vs}
V_\text{pi} = A_\text{pi} w  - A \int_{-\infty}^{0} f(h,t) h \text{d}h \text{  , } \\
V_\text{ps} = A_\text{ps} w  - A r_s \int_{0}^{w} f(h,t) h \text{d}h \text{  , } \\
V_\text{bs} = A_\text{bs} w(1-r_s) \text{  , } 
\end{gather}
\end{linenomath*}
where $A$ is the total area of the domain, and $A_\text{pi}$, $A_\text{ps}$, and $A_\text{bs}$ are areas of ponded ice, ponded snow, and bare snow. After some rearrangement, this finally yields the volume of water as
\begin{linenomath*}
\begin{equation} \label{Vw2}
\frac{V_\text{w}}{A} = w ( 1 - r_s (1-p) ) - \bar{h}(-\infty,0) - r_s \bar{h}(0,w)   \text{ , }
\end{equation}
\end{linenomath*}
where we have introduced a shorthand notation, $ \bar{h}(a,b) \equiv \int_a^b f(h,t) h \text{d}h$. To find the rate of change of $w$, we need to find the rates of change of $\bar{h}$. In general, we have
\begin{linenomath*}
\begin{equation} \label{hbardotgeneral}
\dot{ \bar{h}}(a,b) = b \dot{b} f(b,t)  - a \dot{a} f(a,t) + \int_a^b \dot{f}(h,t) h \text{d}h \text{ . }
\end{equation}
\end{linenomath*}
Applying this rule along with Eq. \ref{dfdt}, applying partial integration once, and using Eqs. \ref{pandpi}, \ref{pandpi_p}, and \ref{pdot1}, we get
\begin{linenomath*}
\begin{gather}\label{hbardot}
\dot{\bar{h}}(-\infty,0) = \dot{h}_\text{pi} p_i \text{  , } \\
\dot{\bar{h}}(0,w) = w \dot{p} + \dot{h}_\text{ps}(p-p_i) \text{  , }
\end{gather}
\end{linenomath*}
where $\dot{h}_\text{pi}$ and $\dot{h}_\text{ps}$ are melt rates of ponded ice and ponded snow. Therefore, we find the rate of change of water volume expressed in terms of the rate of change of water level, coverage fractions and melt rates of different regions of the ice as
\begin{linenomath*}
\begin{equation} \label{dVdt1}
\frac{\dot{V}_\text{w}}{A} = \dot{w}(1-r_s(1-p)) - \dot{h}_\text{pi} p_i - r_s \dot{h}_\text{ps} (p-p_i)  \text{ . }
\end{equation}
\end{linenomath*}
On the other hand, we can express the meltwater production rate using the rates of melt of different regions of the ice
\begin{linenomath*}
\begin{equation} \label{dVdt2}
\frac{\dot{V}_\text{w}}{A} = - \frac{1}{\rho_w} \Big(  \rho_s\dot{h}_s(1-p) + \rho_s \dot{h}_\text{ps} (p-p_i) + \rho_i p_i \dot{h}_\text{pi} \Big) \text{ . }
\end{equation}
\end{linenomath*}
Finally, using Eqs. \ref{dVdt1} and \ref{dVdt2}, we can express the rate of change of the water level as
\begin{linenomath*}
\begin{equation} \label{dwdt}
\dot{w} = - \frac{r_i r_s (1-p) \dot{h}_s - (1-r_i) \big[ r_s p \dot{h}_\text{ps} + p_i (\dot{h}_\text{pi} - r_s \dot{h}_\text{ps})  \big]}{1-r_s(1-p)}\text{ . }
\end{equation}
\end{linenomath*}
This, together with Eq. \ref{pdot2}, represents a solution to pond coverage evolution that makes no approximations under the assumptions stated at the beginning of this section. However, this system of equations is not closed since it still depends on the fraction of ponded ice, $p_i$. So to have a complete solution, Eqs. \ref{pdot2} and \ref{dwdt} would have to be supplemented with an equation for the evolution of $p_i$, which could be achieved by evolving the distribution $f(h,t)$ according to Eq. \ref{dfdt}. This would, however, complicate matters significantly, and we will now show how this can be avoided with an approximation which detracts very little from the accuracy of the solution. 

Firstly, note that, since ice and water have similar densities, $r_i =0.9 \approx 1$, making the second term in Eq. \ref{dwdt}, proportional to $(1-r_i)$, small compared to the first term. Secondly, the first term is proportional to $(1-p)$ while the second term is proportional to $p$ which means that during initial pond growth, the significance of the second term is additionally diminished. Therefore, the second term only becomes important when $p \sim 1$, which is rarely observed in real ponds. By simply neglecting the second term, we get Eqs. 17 and 18 discussed in section VI of the main text
\begin{linenomath*}
\begin{eqnarray} 
&\dot{w} \approx -\frac{r_i r_s (1-p)}{1 - r_s (1 - p)}  \dot{h}_s \text{  , } \label{stageIsimple_supp}\\
&\dot{p}  = f_\Gamma(w - \dot{h}_s t)  \Big( \dot{w} -  \dot{h}_s\Big)  \label{stageIsimple_p_supp} \text{  , }
\end{eqnarray} 
\end{linenomath*}
We show in section \ref{sec:test_pond} that this approximation is very good at the beginning of the melt season and shows a slight deviation when the pond coverage becomes high enough (Fig. \ref{fig:stageIcomparison}). 

\subsubsection{Including pond drainage}

We now note how to include limited drainage during stage I to get Eq. 19 of the main text. If drainage of $Q$ cm per day occurs, this will modify the water balance equation, Eq. \ref{dVdt2}. Simply adding a term $-Q$ to $\frac{\dot{V}_\text{w}}{A}$ in Eq. \ref{dVdt2} and following through the same steps as before yields the correct equation. In this case, the water level equation after neglecting the term proportional to $p_i$ becomes
\begin{linenomath*}
\begin{equation} \label{dwdt2}
\dot{w} = - \frac{r_i r_s (1-p) \dot{h}_s + Q}{1-r_s(1-p)}\text{ , }
\end{equation}
\end{linenomath*}
as discussed in section VI of the main text. $Q$ can be made a function of pond coverage, such that ponds only drain above the percolation threshold. As was also discussed in section VI of the main text, drainage makes it possible for the water level to decrease. This can lead to exposing regions of bare ice and topography that was altered by differential melting, which makes it impossible to use the height distribution $f_\Gamma$. To avoid this, it is sufficient that $\dot{w} -  \dot{h}_s > 0$ and that $w > 0$.  

\subsubsection{Invariants of Eqs. 17 and 18 of the main text}

Finally, we will explain the remark we made in section VI of the main text that the pond evolution is unchanged by increasing the snow melt rate and the volume of snow by the same factor. That is to say, Eqs. 17 and 18 of the main text are invariant under $\dot{h}_s \rightarrow a\dot{h}_s$ and $h(\mathbf{x}) \rightarrow ah(\mathbf{x})$. First, we note that the rate of change of water level, $\dot{w}$, is proportional to the snow melt rate, $\dot{h}_s$, so that $\dot{w} \rightarrow a \dot{w}$ when $\dot{h}_s \rightarrow a\dot{h}_s$, assuming that the pond coverage does not change. From here and using Eq. \ref{pdot2}, it follows that $\dot{p} \rightarrow a f_\Gamma\big( a(w - \dot{h}_s t) \big)\Big( \dot{w} -  \dot{h}_s\Big)$ when $\dot{h}_s \rightarrow a\dot{h}_s$. Second, from the definition of the Gamma distribution (Eq. 9 of the main text), we can see that when the scale, $h_0$, transforms as $h_0 \rightarrow a h_0$, the gamma distribution transforms as $f_\Gamma(x) \rightarrow \frac{1}{a}f_\Gamma(x/a)$. Equation 10 of the main text relates $h_0$ to the mean and the variance of the surface height as $h_0 = \sigma^2(h)/\langle h \rangle$, from which it follows that changing the surface height as $h(\mathbf{x}) \rightarrow ah(\mathbf{x})$ leads to $h_0 \rightarrow a h_0$. Combining the transformation properties of the gamma distribution under $h(\mathbf{x}) \rightarrow ah(\mathbf{x})$ and the transformation properties of the water level under $\dot{h}_s \rightarrow a\dot{h}_s$, we find that the pond coverage evolution described by $\dot{p}$ is unchanged. 

\subsection{Second-order approximateion}

The approximation we made to get Eq. \ref{stageIsimple_supp} can be improved by noting that the term proportional to $p_i$ in Eq. \ref{dwdt} is a difference between melt rates of different types of ponded surface, $\dot{h}_\text{pi} - r_s \dot{h}_\text{ps}$. Expressing this difference in terms of heat fluxes, we can see it is equal to $(\alpha_\text{ps} - \alpha_\text{pi})\frac{F_\text{sol}}{l_m \rho_i}$, where $\alpha_\text{ps}$ and $\alpha_\text{pi}$ are the albedos of ponded snow and ponded ice. Therefore, if ponded snow and ponded ice have similar albedos, this difference is small. For this reason, it is justified to neglect the term proportional to $p_i$ in Eq. \ref{dwdt}. Under this approximation, we get a closed system of equations 
\begin{linenomath*}
\begin{gather} \label{stageIsecond}
\dot{w} = - \frac{r_i r_s (1-p) \dot{h}_s - (1-r_i) r_s p \dot{h}_\text{ps}}{1-r_s(1-p)}\text{ , }\\
\dot{p}  = f_\Gamma(w - \dot{h}_s t)  \Big( \dot{w} -  \dot{h}_s\Big)  \text{  . } \label{stageIsecond_p}
\end{gather}
\end{linenomath*}
In the next subsection, we show that these equations are nearly indistinguishable from the full 2d model. Equations \ref{stageIsecond} and \ref{stageIsecond_p} are as simple to solve as Eqs. \ref{stageIsimple_supp} and \ref{stageIsimple_p_supp}, but we chose to focus on Eqs. \ref{stageIsimple_supp} and \ref{stageIsimple_p_supp} in the main text since they highlight the dominant role of snow melt rate. We can add drainage in this apporximation in the same way as in the previous subsection 
\begin{linenomath*}
\begin{equation} \label{dwdt2}
\dot{w} = - \frac{r_i r_s (1-p) \dot{h}_s - (1-r_i) r_s p \dot{h}_\text{ps} + Q}{1-r_s(1-p)}\text{ , }
\end{equation}
\end{linenomath*}

Including additional details is also possible within this framework. For example, if melt rates change with time all of the derived evolution equations can still be used, except that the term $\dot{h}_s t$ entering $f_\Gamma$ in Eq. \ref{pdot2} should be replaced with $\langle \dot{h}_s \rangle t$, where $\langle \dot{h}_s \rangle$ means mean snow melt rate up to time $t$.

\subsection{Testing the analytic model}\label{sec:test_pond}

\begin{figure}
\centering
\includegraphics[width=0.6\linewidth]{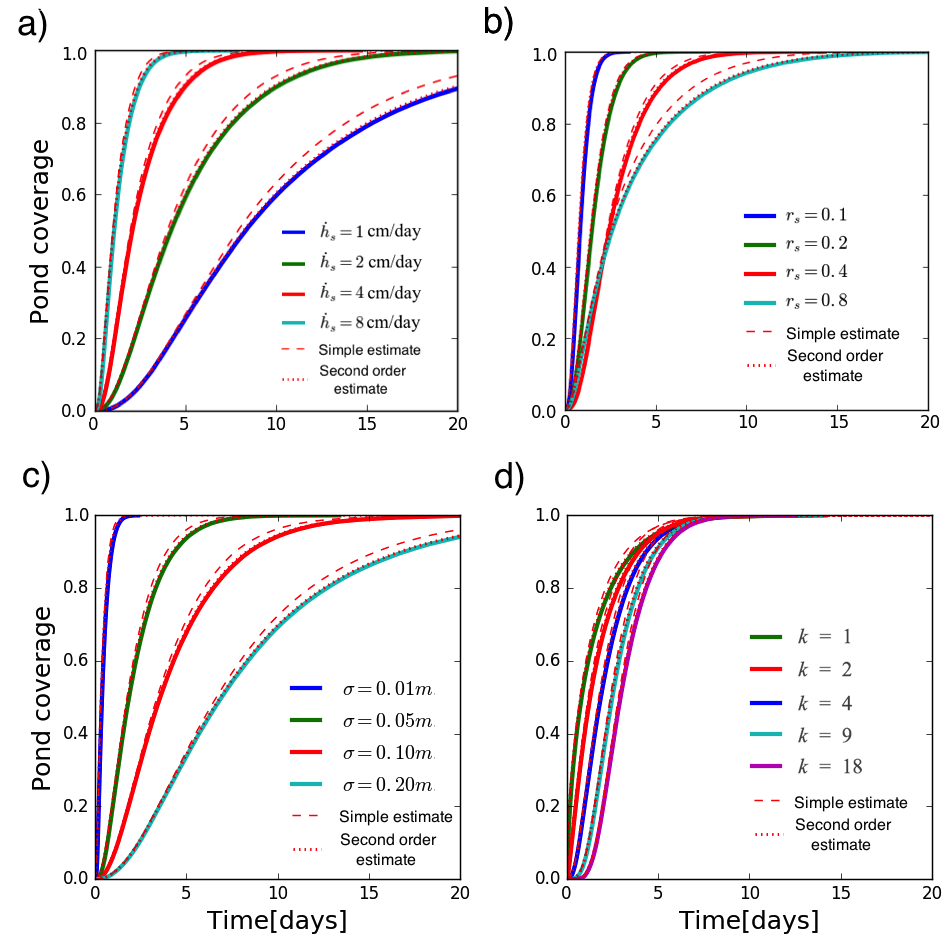}
\caption{Pond coverage evolution during stage I for a variety of model parameters. The solid lines are solutions to the full stage I 2d model on a ``snow dune'' topography for different model parameters, the red dashed lines are solutions to Eqs. \ref{stageIsimple_supp} and \ref{stageIsimple_p_supp}, while the red dotted lines are solutions to Eqs. \ref{stageIsecond} and \ref{stageIsecond_p}. As in Fig. 5 of the main text, in each panel, we change one parameter, while we keep the others fixed at default values $\dot{h}_s = 4 \text{ cm day}^{-1}$, $r_s = 0.4$, $\sigma(h) = 0.05\text{ m}$, and $k = 4$ a) Pond coverage evolution for different snow melt rates, $\dot{h}_s$. b) Pond coverage evolution for different snow to ice density ratios, $r_s$. c) Pond coverage evolution on topographies with different snow-depth standard deviations, $\sigma(h)$. d) Pond coverage evolution on topographies with different shape parameters, $k$. }
\label{fig:stageIcomparison}
\end{figure}

The assumptions stated at the beginning define a model that can be solved numerically on the full 2d topography with all of the details of the topography included. In this case, we initiate the model by generating a ``snow dune'' topography with a desired mean and variance. Then, at each time step, we find the water level, use it to find regions of ponded ice and ponded snow, update the topography by melting different regions of the surface at their prescribed rates, find the volume of water generated, and, finally, update the water level again. We show the results of these simulations as colored lines in Fig. \ref{fig:stageIcomparison}.

In Figs. \ref{fig:stageIcomparison}, we compare the simple estimate, Eqs. \ref{stageIsimple_supp} and \ref{stageIsimple_p_supp} (red dashed lines), and the second order estimate, Eqs \ref{stageIsecond} and \ref{stageIsecond_p} (red dotted lines), with the solutions to the full 2d model (colored lines). We can see that the simple estimate is a good approximation to the full 2d model in the beginning and shows slight discrepancy at large $p$, as expected. We can also see that the second-order approximation is nearly indistinguishable from the full 2d solutions. 

\newpage


\bibliography{mybibliogJGR.bib}